\documentclass[conference]{IEEEtran}
\usepackage{mathptmx}
\pdfoutput=1
\usepackage[normalem]{ulem}

\usepackage{color}
\usepackage{cite}
\usepackage{threeparttable}
\usepackage{float}
\usepackage{amsmath,amssymb,amsfonts}
\usepackage{graphicx}
\usepackage{textcomp}
\usepackage{multirow}
\usepackage{colortbl}
\usepackage[ruled]{algorithm2e}
\usepackage{diagbox}
\usepackage{mathrsfs}
\usepackage{booktabs}
\usepackage{hhline}
\usepackage{array,ragged2e}
\usepackage{tabularx}
\usepackage{algorithmic}
\usepackage{xcolor}
\usepackage{lipsum}

\usepackage{verbatim}
\usepackage{enumitem}
\usepackage[colorinlistoftodos,prependcaption,textsize=small]{todonotes}
\usepackage[ruled]{algorithm2e}
\usepackage{multicol,multicap}

\usepackage{subfig}
\usepackage{listings}
\usepackage{fancyhdr} 

\usepackage{soul}

\definecolor{bg}{rgb}{0.96,0.96,0.96}

\newcommand{\Fig}[1]{Fig.~\ref{#1}}

\newcommand{\Tbl}[1]{Tbl.~\ref{#1}}
\newcommand{\Sec}[1]{Sec.~\ref{#1}}

\renewcommand{\paragraph}[1]{\vspace*{.1cm}\noindent\textbf{#1}\hspace*{.1cm}}

\newcommand{\specialcell}[2][c]{\begin{tabular}[#1]{@{}c@{}}#2\end{tabular}}

\newcommand{\im}{\texttt{im2col}}
\newcommand{\conv}{\texttt{CONV}}

\newcommand*\circledwhite[1]{\tikz[baseline=(char.base)]{
            \node[shape=circle,draw,inner sep=0pt] (char) {#1};}}
            
\def\BibTeX{{\rm B\kern-.05em{\sc i\kern-.025em b}\kern-.08em
    T\kern-.1667em\lower.7ex\hbox{E}\kern-.125emX}}

\title{\huge Characterizing and Demystifying the Implicit Convolution Algorithm on Commercial Matrix-Multiplication Accelerators}

\author{
  \IEEEauthorblockN{
  \small Yangjie Zhou$^{1,2}$, Mengtian Yang$^1$, Cong Guo$^{1,2}$, Jingwen Leng$^{1,2}$, Yun Liang$^{3,2}$, Quan Chen$^{1,2}$, Minyi Guo$^{1,2}$, Yuhao Zhu$^4$\\
$^1$\textit{Shanghai Jiao Tong University}, 
$^2$\textit{Shanghai Qi Zhi Institute}, 
$^3$\textit{Peking University},
$^4$\textit{University of Rochester}
}
}

\begin{document}

\maketitle
\pagestyle{empty}

\begin{abstract}

Many of today's deep neural network accelerators, e.g., Google's TPU and NVIDIA's tensor core, are built around accelerating the general matrix multiplication (i.e., GEMM). 
However, supporting convolution on GEMM-based accelerators is not trivial. 
The naive method explicitly lowers the convolution to GEMM, commonly known as \im{}, which introduces significant performance and memory overhead.
Existing implicit \im{} algorithms require unscalable hardware and are inefficient in supporting important convolution variants such as strided convolution.
In this paper, we propose a memory-efficient and hardware-friendly implicit \im{} algorithm used by Google's TPU, which dynamically converts a convolution into a GEMM with practically zero performance and memory overhead, fully unleashing the power of GEMM engines. Through comprehensive experimental results, we quantitatively argue that this algorithm has been adopted in commercial closed-source platforms, and we are the first to describe its high-level idea and implementation details. 
Finally, we show that our algorithm can also be generally applied to Nvidia's Tensor Cores (TC), matching and out-performing the measured performance on TCs.

\end{abstract}
\section{Introduction}

The recent development of convolutional neural network (CNN) models~\cite{lecun2015deep} has lead to its wide adoption in many fields such as autonomous driving~\cite{levinson2011towards, janai2020computer, yurtsever2020survey} and natural language processing~\cite{hochreiter1997long, devlin2018bert, chowdhury2003natural}. Yet, many commercial neural network accelerators, such as Google's TPU~\cite{tpu_paper}, NVIDIA's Tensor Cores (TCs) since the Volta architecture~\cite{v100}, Habana Gaudi~\cite{medina2020habana}, and Intel's NNP-T~\cite{yang2019deep}, choose the general matrix-matrix multiplication (GEMM) as the basic computation primitive.

It is non-trivial to support CNNs on the GEMM-specialized accelerator.
Many recent works~\cite{kim2020duplo, sparten, zhu2019sparse} make the assumption of explicit \im{} (image-to-column) algorithm, which lowers the convolution to a matrix multiplication via input transformation.
The naive approach performs an \textit{explicit} \im{} transformation to prepare the lowered feature map in the form of the expanded matrix.
As such, this matrix can be consumed directly by the GEMM engine without any hardware modifications. 
This explicit \im{} transformation leads to significant performance and memory overheads.

Commercial GPUs adopt the \textit{implicit} \im{} algorithm~\cite{li2016performance} to avoid the performance and memory overheads in the explicit algorithm.
However, the exact implicit algorithm is not published and it is unclear how to implement it on GEMM-based accelerators like TPUs.
In this work, we study the only described implicit \im{} method in the public domain~\cite{lym2019delta}.
We find that it requires an unscalable hardware design (heavily-banked memory with a large crossbar) for porting to the TPU, and is also inefficient in executing common \conv{} variants such as strided and dilated \conv{}~\cite{chen2017deeplab}.

In this paper, we demystify a hardware-friendly and memory-efficient implicit \im{} algorithm used by the TPU, which dynamically converts a convolution into a GEMM with practically zero performance and memory overhead, fully unleashing GEMM engines' power. 
Such an implicit algorithm leverages the associativity and commutativity in convolution, and requires the memory layout and tiling strategy optimization.
As such, the GEMM engine in the TPU is served with data from a simple single-bank memory while allowing off-chip memory access and computation to be fully overlapped.

We develop a configurable cycle-level TPU simulator for performance evaluation. Our simulator is validated against the cloud TPUv2 measurement and has an average error rate of less than 5\%.
We plan to make the simulator open-source to encourage more study. 

In addition to the TPU, we also show that our implicit \im{} can also be applied to the TCs on the GPU. The challenge is to maximally utilize the many TCs on the GPU. To that end, we devise a blocked version of our \im{} algorithm. We exploit tile reordering to avoid stalling from off-chip memory accesses.
We implement and evaluate our algorithm on a V100 GPU. 
The difference in performance between us and cuDNN is 1\% at batch size of 8.

To our best knowledge, we are the first in the public domain to introduce a generic implicit \im{} algorithm that is implemented on both a systolic array and the TCs. We do not know (nor claim) whether (part of) our design is implemented in TPU or the TCs in Nvidia's GPUs. That said, we present our educated guess as to what part of our design is likely implemented in the TPU and/or the TCs, and why our design achieves higher performance than the proprietary TPU and GPU designs in certain scenarios.

In summary, the paper makes the following contributions:

\begin{itemize}[leftmargin=*]
	\item We quantify the performance and memory overhead of explicit \im{} method over implicit \im{} method.
	\item To our best knowledge, we study the first open, public design of implicit \im{}, which is generally applicable to GPUs and systolic array-like accelerators (e.g., TPUs) with zero memory and near-zero performance overhead.
	\item We implement two concrete instances of the above implicit \im{} method on the commodity TPUs and GPUs, via simulation and software implementation respectively. We show that our methods are on-par with and sometimes even better than the vendor's proprietary implementations.
	\item All the artifacts are made available. We hope our design can shed some light upon, and identify potential room for improvement of the proprietary designs.
\end{itemize}

We organize the paper as follows. \Sec{sec:motivation} introduces the background on \im{} and quantifies the inefficiencies of existing \im{} methods. \Sec{sec:algo} describes our \im{} methods. We then describe how to implement and optimize our \im{} on the TPU (\Sec{sec:tpu}) and the TCs on the GPU (\Sec{sec:gpu}). After the experimental methodology (\Sec{sec:methodology}), we evaluate our \im{} designs on the TPU and the GPU (\Sec{sec:eval}). \Sec{sec:related} describes the related work and \Sec{sec:conclusion} concludes the paper.

\section{Motivation}
\label{sec:motivation}

This section first introduces the background on \im{} (\Sec{sec:motivation:background}). We then quantitatively demonstrate that explicit \im{} is both memory inefficient and slow (\Sec{sec:motivation:ex}). However, existing implicit \im{} implementations have inherent limitations and/or are proprietary (\Sec{sec:motivation:limitation}).

\subsection{Background on Im2Col}
\label{sec:motivation:background}

Many commercial hardware accelerators for deep neural network (DNN), such as Google's TPU~\cite{tpu_paper}, NVIDIA's Tensor Cores since the Volta architecture~\cite{v100}, Habana's Gaudi~\cite{medina2020habana}, and Intel's NNP-T~\cite{yang2019deep}, choose the general matrix-matrix multiplication (GEMM) as the basic computation primitive.
The reason is that most of the operations in DNN models, including the fully connected and convolutional layer, can be directly mapped or lowered to GEMMs.

\begin{figure}[t]
    \centering 
    \includegraphics[width=\columnwidth]{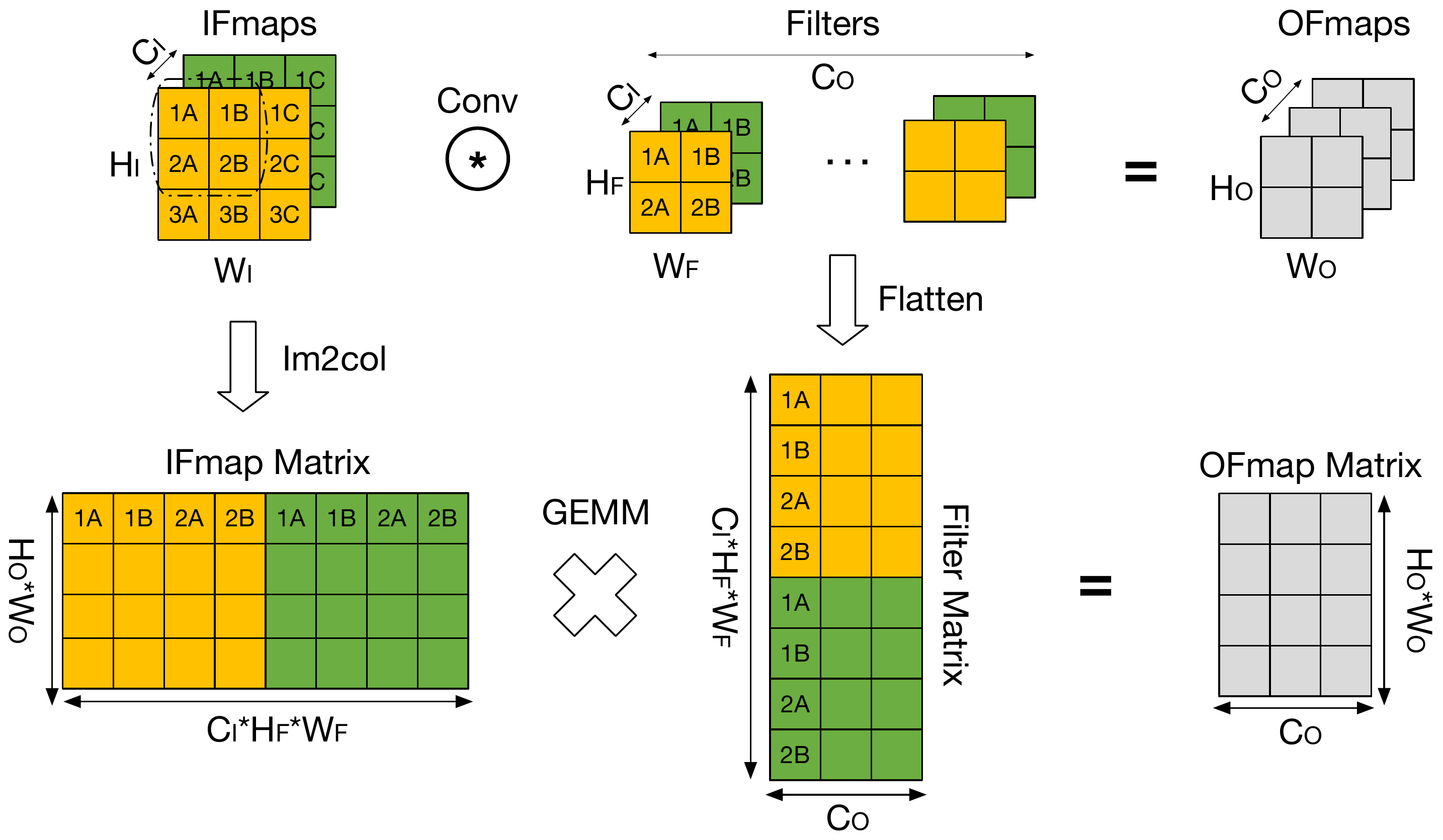}
    \caption{\small Illustration of the \im{} algorithm that converts a \conv{} layer to a GEMM operation. For simplicity, this examples assumes no padding in the IFMap.}
    \label{fig:im2col}
    \vspace*{-0.3cm} 
\end{figure}

Convolution is still an important and fundamental workload, e.g., Google has acknowledged that the portion of CNNs in its data centers increases from 5\% to 24\%~\cite{tpu_cacm}.
The upper part of \Fig{fig:im2col} shows the details of a \conv{} layer.
It takes in a number of $C_I$ input feature maps (or input channels), each sized of $H_I\times W_I$. Every input feature map is convolved by a sliding kernel (or weight) size of $H_F\times W_F$ to calculate one pixel in the output feature map. A total of $C_O$ feature maps (or output channels) will be generated as output to the next layer.
In many settings (e.g., training), a batch of N inputs can be executed in parallel to amortize the cost of weight accesses.

However, lowering DNN operations to GEMM is not automatic. \im{} (image-to-column) is the de facto algorithm used to lower a convolution to a GEMM~\cite{chellapilla2006high, chetlur2014cudnn}. 
The bottom of \Fig{fig:im2col} shows an example of this transformation.
The ($H_I\times W_I\times C_I$) IFMap is first expanded into a ($H_O W_O \times H_F W_F C_I$) matrix, which we call the \textit{lowered} feature matrix. Each row in the lowered matrix corresponds to the receptive field of an element in the OFMap, as \Fig{fig:im2col} shows. The filters are then flattened to a matrix with the size of ($C_I H_F W_F \times C_O)$.

\subsection{The Need for Implicit Im2Col}
\label{sec:motivation:ex}

The naive approach performs an explicit \im{} transformation to prepare the lowered feature map before the latter is consumed by the GEMM engine (e.g., the Tensor Cores in Nvidia's GPUs or the TPU). This explicit \im{} transformation leads to significant performance and memory overheads over the GEMM computation itself.

Our measurements show that neither Nvidia's GPUs nor the TPU uses explicit \im{} (although the option is available on Nvidia's GPUs), presumably because of the high overhead. Instead, their proprietary implementations, which we call the \textit{implicit} \im{}, show little overhead. We quantitatively demonstrate the inefficiencies of explicit \im{} to motivate implicit \im{}. To our best knowledge, no such analysis is available for TPU, nor other GEMM accelerators.

\begin{table}[b]
    \caption{Memory usage (MB) breakdown for executing different CNNs with the explicit \im{} on V100 GPU.}
    \label{tab:workspace}
    \resizebox{\linewidth}{!}{
    \begin{tabular}{|c|ccccc|}
        \hline
        & AlexNet& ResNet& VGG16& YOLO& DesNet\\
        \hline
        IFmaps& 1.39& 34.55& 34.65& 530.56& 1196.48\\
        Lower IFmaps& 14.57& 81.11& 311.80& 869.50& 5641.70\\
        \hline
    \end{tabular}}
    \end{table}

\paragraph{Memory Overhead} The lowered feature matrix from the \im{} transformation takes up to $H_F\times W_F$ times more memory than the original feature map, because the overlapped receptive fields generate duplicated data.

We use GPU as a case study to demonstrate the overhead. \Tbl{tab:workspace} compares the memory required for storing the lowered input matrix in the explicit \im{} method across a range of different models. As a reference, we also show the original size of the input feature (IFMap). The data is measured on a V100 GPU using the explicit \im{} APIs from the cuDNN library~\cite{chetlur2014cudnn}. The additional storage requirement is generally $1.5\times - 10\times$ of the input feature maps.

\begin{figure}[t]
    \vspace*{-0.3cm}
    \centering
    \subfloat[NVIDIA V100 GPU.]
    {
      \includegraphics[trim=0 0 0 0, clip, width=0.49\linewidth]{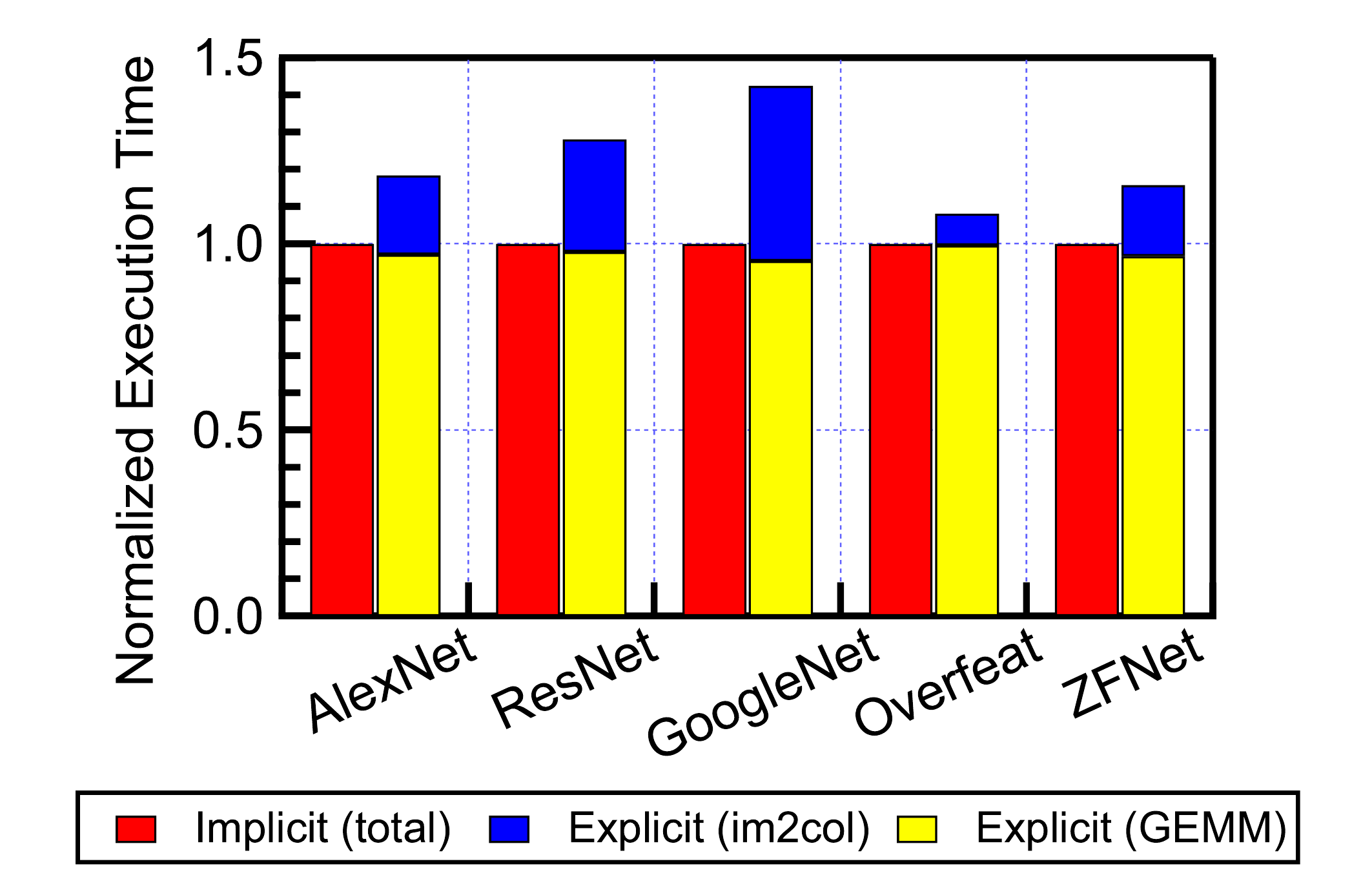}
   	  \label{fig:gpu_64_time}
    }%
    \subfloat[Google TPU-v2.]
    {
      \includegraphics[trim=0 0 0 0, clip, width=0.49\linewidth]{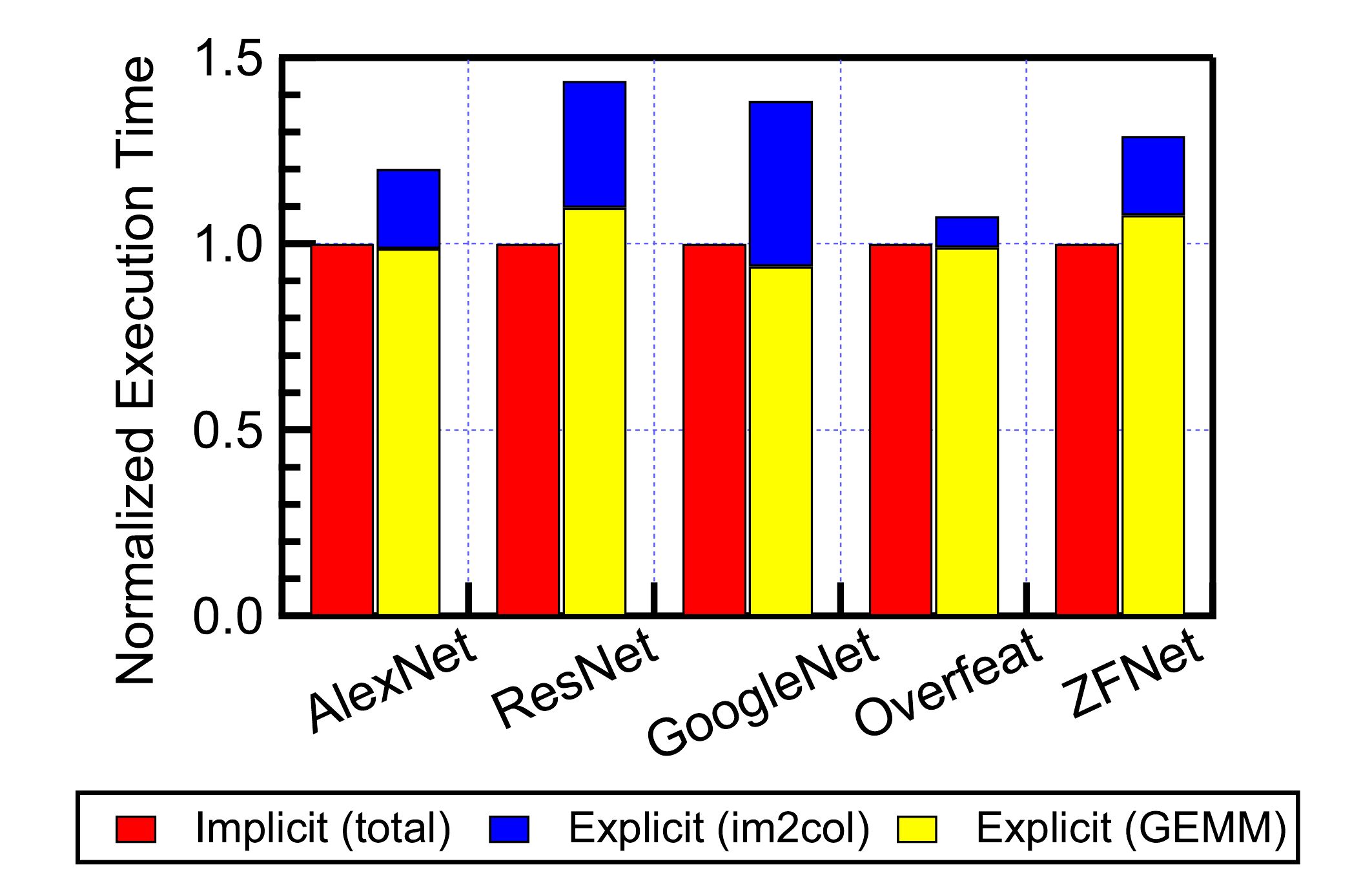}
      \label{fig:tpu_64_time}%
    }
    \vspace*{-0.2cm}
    \caption{\small The execution time comparison of explicit and implicit \im{} methods for convolutional layer on V100 GPU and TPU-v2. We use a batch size of 64 for all CNNs.}
    \label{fig:cudnn_time}
    \vspace*{-0.4cm}
\end{figure}

\paragraph{Performance Overhead} Explicit \im{} also introduces significant performance overhead. To demonstrate this, we measure the GPU execution time of both the implicit and explicit versions using their corresponding cuDNN APIs. The stacked bars in \Fig{fig:gpu_64_time} break down the execution time on the V100 GPU into the GEMM time and the explicit \im{} transformation time (batch size 64). The execution time is normalized to the execution time of the implicit method.

We find that explicit \im{} is 28\% slower than the implicit approach on average.
Critically, the GEMM time in the explicit method is almost identical to the implicit method. This suggests that the implicit method has near-zero performance overhead: all its time is spent on GEMM.
In other words, the explicit method introduces a performance overhead of about 26\% compared to the GEMM on GPU.

The conclusion holds on the cloud TPU-v2 platform~\cite{tpucloud}. \Fig{fig:tpu_64_time} compares the execution time of both methods on TPU-v2. Note that the TPU does not provide the explicit option; we thus mimic the behavior of explicit \im{} as if it was to be used on the TPU by combining the GEMM time on TPU and the time of explicit \im{}, which is estimated by using the GPU results. This strategy provides a performance \textit{lower bound} for the explicit \im{} if it was supported on TPU, because we omit the overhead of transmitting the lowered matrix to the TPU.

On average, the explicit \im{} transformation introduces 26\% overhead. The explicit method is 23\% slower than the implicit method, whose execution time is roughly the same as the GEMM time in the explicit method, indicating little performance overhead of \im{} in the implicit method.

\begin{figure}[t]
    \centering
    \includegraphics[width=\linewidth]{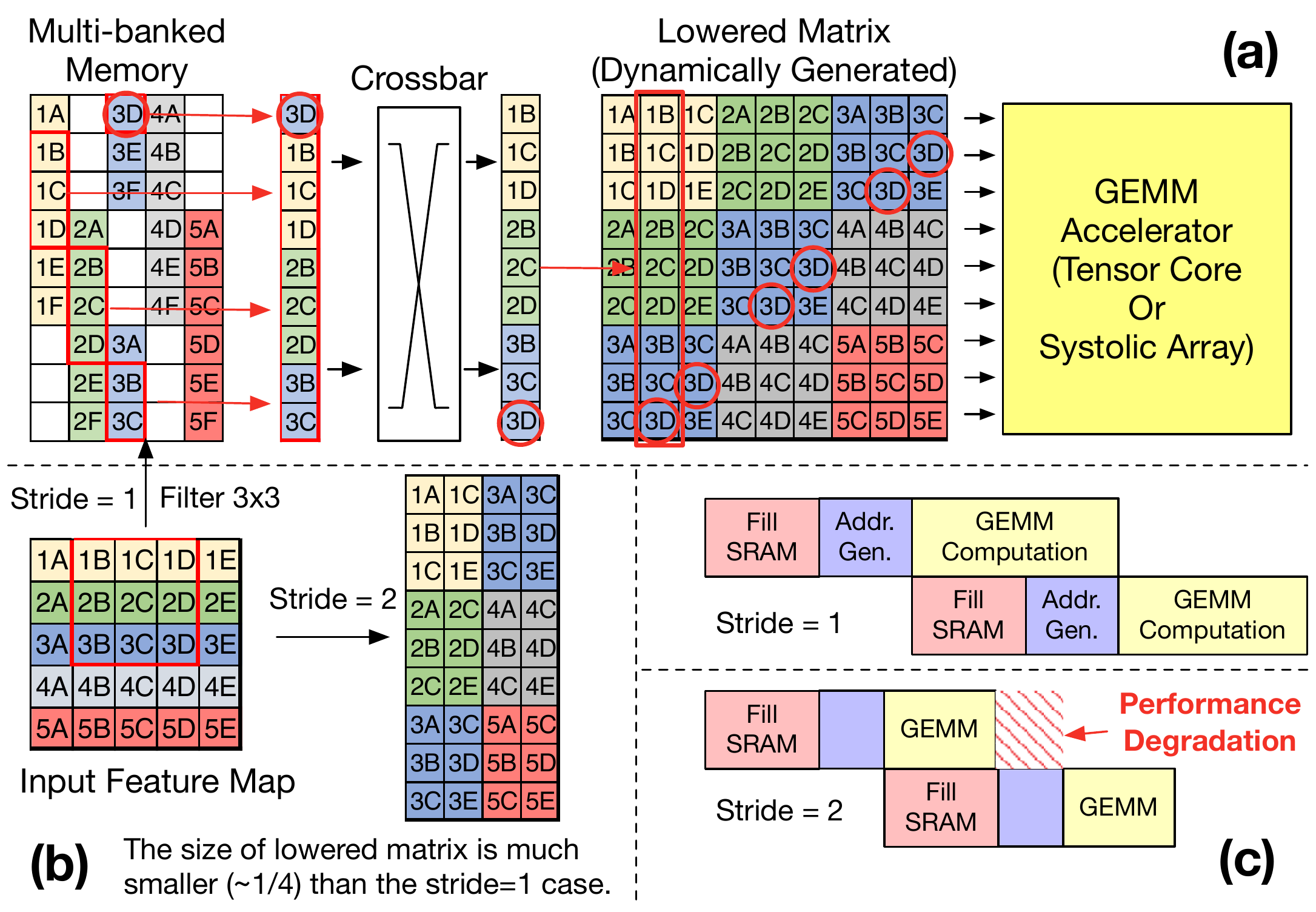}
    \vspace*{-0.2cm}
    \caption{\small (a) The existing implicit \im{} design~\cite{lym2019delta} dynamically forms the lowered IFMap, assuming a stride of 1. It has little performance overhead, but requires a multi-banked on-chip memory and a large crossbar between the memory and the GEMM engine.
    (b)\&(c) The $stride=2$ causes throughput degradation: the latency to fill in the on-chip memory could not be hidden by the GEMM latency, which is about 1/4 of the $stride=1$ case.}
    \label{fig:implicit_im2col_challenge}
    \vspace*{-0.5cm}
\end{figure}

\subsection{Limitations of Existing Implicit Approach}
\label{sec:motivation:limitation}

Characterization results suggest that both GPU and TPU use some forms of implicit \im{} method. However, their implementations are proprietary. 
We describe a prior academic effort based on Lym et al.~\cite{lym2019delta}, a conceptually clear design to support implicit \im{} for GPUs' CUDA Core. We migrate this design to Tensor Core. The difference between the two computational patterns is that for the CUDA core, the warp-level GEMM is computed via outer product, while the TensorCore computes GEMM via inner product~\cite{raihan2019modeling}.
We do not claim (nor know) whether the design by Lym et al. is the same as that in GPUs, but we show that today's GPUs suffer from some of the similar inefficiencies to that of Lym et al. Thus, Lym et al. would provide a sensible design to understand the inefficiencies in existing implicit \im{} implementations.

The basic idea of Lym et al.~\cite{lym2019delta} is to use a flexible on-chip memory structure to dynamically form the lowered feature matrix before the latter is fed into the compute engine. \Fig{fig:implicit_im2col_challenge} illustrates the details. The input feature map (IFMap) is stored in the on-chip SRAM (e.g., the shared memory in the GPU). Each element in the IFMap is dynamically routed to the correct PE in the GEMM engine, effectively forming the lowered IFMap (middle matrix) at run time.

The advantage of Lym et al. is two-fold. First, the lowered IFMap does not require additional storage because it is dynamically formed and immediately consumed. 
Second, it can sustain the full throughput of the GEMM engine. This is because filling the on-chip memory for the next block/tile (through accessing DRAM) and GEMM computation can be overlapped, as shown in the lower right panel in \Fig{fig:implicit_im2col_challenge}.

However, the hardware design by Lym et al. (and today's GPUs) does not generally scale to other forms of GEMM accelerator and/or incurs significant performance overhead for common convolution variants such as strided and deformable convolution~\cite{deformableconv}. Let us elaborate below.

\paragraph{Unscalable Hardware}
The main requirement of prior work is a multi-banked SRAM with a large crossbar, which routes elements in the IFMap (stored in the SRAM) to the correct PEs in the GEMM engine at each cycle. 
Consider the second column in the lowered matrix in \Fig{fig:implicit_im2col_challenge}. All 9 elements that enter the GEMM engine in one cycle have to come from different banks of the SRAM in order to not stall the GEMM engine. The bank conflict can be avoided by carefully laying out the IFMap elements in the SRAM offline~\cite{lym2019delta}.

Critically, each element in the SRAM needs to be mapped to different PEs at different cycles, entailing a crossbar. E.g., the element \circledwhite{\small 3D} at the current cycle maps to the PE in the last row, but will map to the second last row next cycle.

Lym et al. made an astute observation that modern GPUs naturally provide such a multi-banked SRAM (i.e., the shared memory with 32 banks) and a crossbar (i.e., the $32 \times 32$ crossbar for shuffling data within a warp) in each SM. Therefore, this implicit \im{} design introduces little additional hardware on top of existing GPU hardware.

However, this design is unscalable, because the size of the crossbar and the number of banks in the SRAM would have to scale proportionally to the PE array size in the GEMM engine. For instance, TPUv1~\cite{tpu_paper} uses a $256 \times 256$ PE array, requiring a $256 \times 256$ crossbar and a 256-bank SRAM. The crossbar area and power increase quadratically with respect to the number of ports~\cite{kilonoc}, and a large number of memory banks also degrades the area and power efficiency~\cite{645796}.

\begin{figure}[t]
    \vspace*{-0.2cm}
    \hspace*{-0.3cm}
    \centering
    \subfloat[NVIDIA V100 GPU.]
    {
      \includegraphics[width=0.5\linewidth]{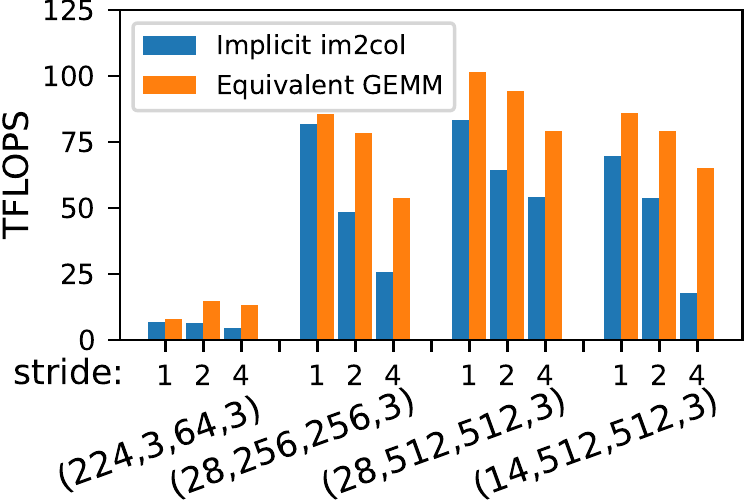}
   	  \label{subfig:stride_conv_gpu}
    }
	%
    \subfloat[Google TPU-v2.]
    {
      \includegraphics[width=0.5\linewidth]{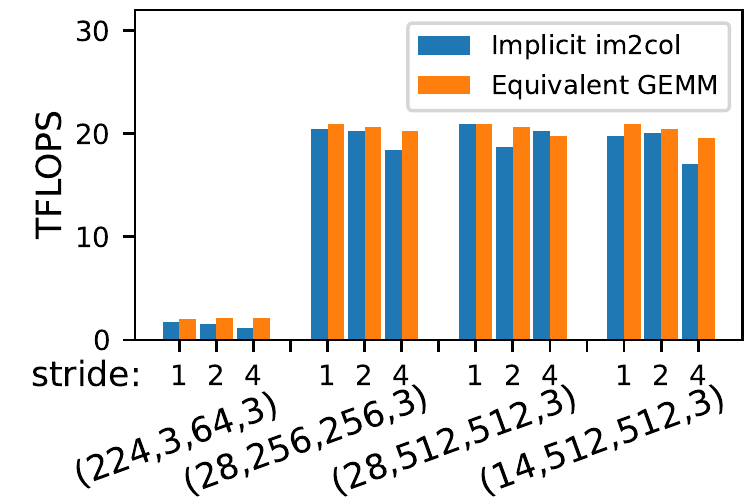}
   	  \label{subfig:stride_conv_tpu}%
    }
    \caption{\small TFLOPS of implicit \im{} on representative ResNet layers (represented by $W_I, C_I, C_O, W_F$) under different strides. As a reference, we also show the TFLOPS of GEMM only. GPU performance (a) significantly degrades with larger strides, whereas the TPU (b) is insensitive to the stride.}
    \label{fig:stride_conv}
    \vspace{-0.6cm}
\end{figure}

\paragraph{Supporting \conv{} Variants} The performance of the existing implicit \im{} approach degrades significantly for key convolution variants such as the strided convolution, which are commonly used in modern CNNs~\cite{kong2017stride}.

\Fig{subfig:stride_conv_gpu} shows the performance measured in TFLOPS (Tera Floating Point Ops per Second) of several representative layers of ResNet~\cite{he2016deep} under different strides on the Tensor Cores of V100 GPU.
Compared with the stride as 1, the GPU performance drops by 30\% under a stride of 2 and 60\% under a stride of 4.
To understand the performance drop, \Fig{subfig:stride_conv_gpu} also shows the TFLOPS of a GEMM kernel operating on a matrix of the same size as the lowered IFMap. The GEMM's TFLOPS is much higher than that of the implicit \im{} under larger striders, indicating the implicit method becomes severely memory-bound with a greater-than-one stride.

Implicit \im{} becomes memory-bound under larger strides because the GEMM latency (of a tile) reduces; thus, the latency of filling the on-chip SRAM could not be hidden by the GEMM computation.
The bottom half of \Fig{fig:implicit_im2col_challenge} illustrates the reason.
With $stride=1$, the address generation overhead to access the multi-banked memory (\Fig{fig:implicit_im2col_challenge} top) makes the overall performance roughly the same as the equivalent GEMM computation in \Fig{fig:stride_conv}.
This indicates the GEMM tile latency in \Fig{fig:implicit_im2col_challenge} bottom can just overlap with the SRAM filling and address generation process together.
However, with $stride=2$, the size of the lowered matrix is reduced significantly (about 1/4 for the stride of two).
Thus, the GEMM latency is reduced significantly while the SRAM loading time does not change, leading to large performance degradation.

TPU does \textit{not} show the same performance degradation. \Fig{subfig:stride_conv_tpu} shows the results of the same experiment but on TPU. The results demonstrate that TPU performance is insensitive to the stride value.
The difference between the TPU and the GPU suggests that the TPU and the GPU potentially use different implicit \im{} designs --- there is a design space for implicit \im{} that has not been explored in prior work.

\paragraph{Summary} Existing implicit \im{} design in the public domain requires a complicated, unscalable hardware design and is inefficient in handling common convolution variants. We observe that some of these issues exist in the proprietary designs from commercial accelerators as well.

We study a previously unpublished implicit \im{} algorithm that addresses both issues. 
Our real hardware measurement results show that this algorithm is highly possible used in the TPU architecture. 
We also demonstrate its general feasibility and advantage on the tensor core GEMM accelerator.

\begin{figure}[t]
    \centering
    \includegraphics[width=\linewidth]{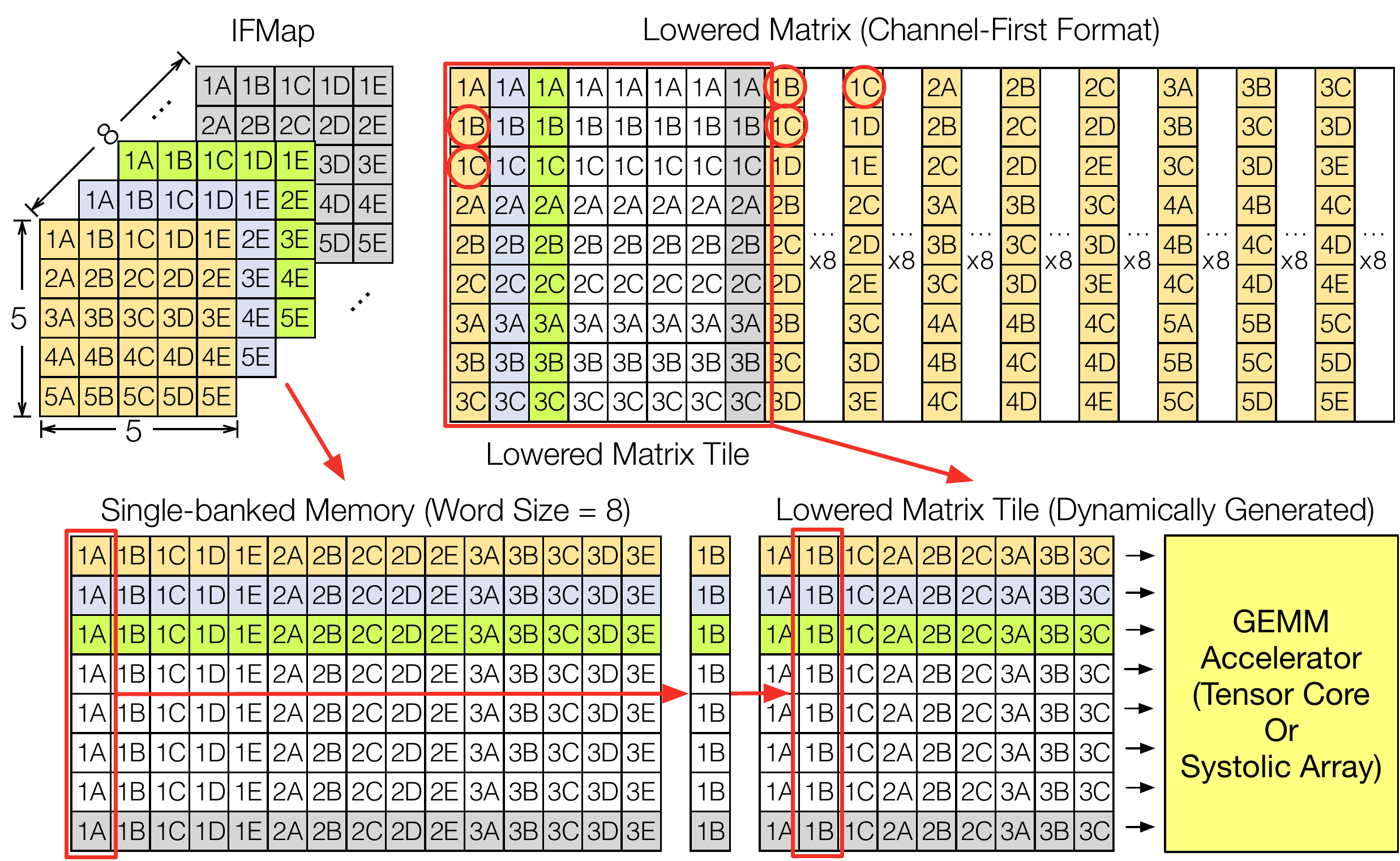}
    \vspace{-0.4cm}
    \caption{\small We store the IFMap with the $HWC$ format in the on-chip SRAM. This allows the GEMM to read data from the SRAM as a wide word rather than from different banks, simplifying the hardware.}
    \label{fig:benefit_c_first_im2col}
     \vspace{-0.5cm}
\end{figure}

\section{Implicit Channel-First Im2col}
\label{sec:algo}

In this section, we explain a powerful \im{} algorithm called channel-first \im{} method, which avoids the two sources of inefficiency associated with today's \im{} algorithm. We first present the basic idea that allows us to use a simple crossbar-free, single-bank on-chip memory to serve IFMap to the GEMM engine (\Sec{sec:method:idea}). We show that the basic idea can be trivially extended to balance the speed of GEMM computation and filling the SRAM, eliminating the inefficiencies in computing \conv{} variants (\Sec{subsec:conv_variant}).

\subsection{Channel-First Im2col Method}
\label{sec:method:idea}

The basic idea of our approach is to layout the IFMap in the on-chip SRAM in such a way that each IFMap element is sent to a deterministic PE. In this way, we can avoid the costly multi-banked SRAM and the associated crossbar.

We will first use a concrete example to describe how such an IFMap layout looks and why such a layout ensures correctness. We will then explain the general principle behind such a layout, which is conceptually nothing more than simply reordering elements in the lowered IFMap matrix.

\begin{figure}[t]
    \centering
    \vspace{-0.2cm}
    \includegraphics[width=\linewidth]{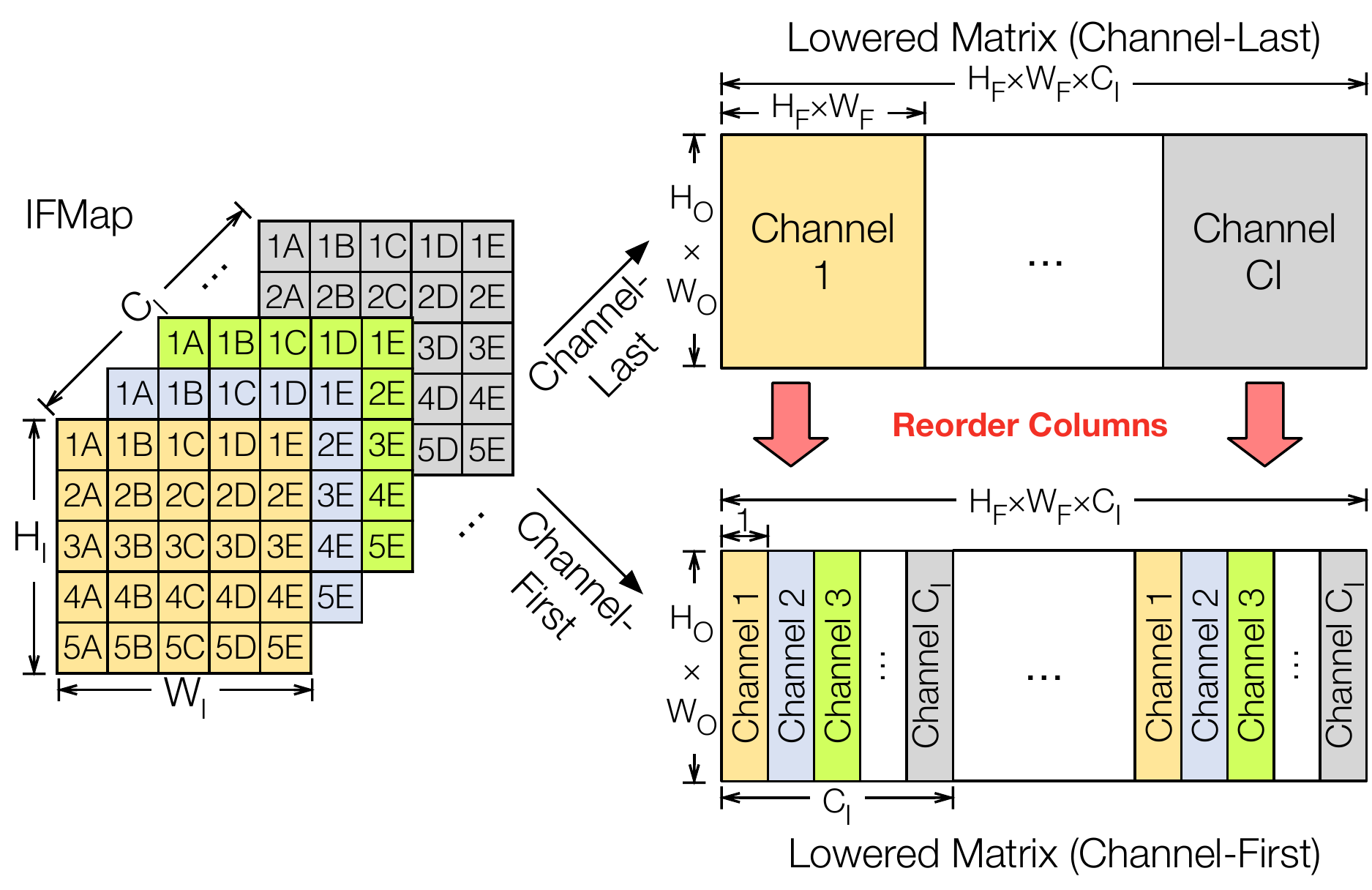}
    \vspace{-0.6cm}
    \caption{\small The channel-last and -first layout for the lowered matrix.}
    \label{fig:im2col_hw_vs_c}
    \vspace{-0.4cm}
\end{figure}

\paragraph{On-chip SRAM Data Layout} 
Assuming an IFMap with 8 channels ($C_I$), each with the dimension $5 \times 5$ ($H_I\times W_I$), \Fig{fig:benefit_c_first_im2col} shows our proposed IFMap layout in the SRAM, where each row is an unrolled vector of a channel in the IFMap. In other words, each column consists of elements of the same position across channels. For instance, the 8 elements in the first column are the elements at position 1A from all 8 channels. Each cycle, one column is read out as a whole word from the SRAM, and fed into the GEMM engine.

We call this is the \textit{Channel-First}, or the HWC layout, since the channel dimension is unrolled first. Note that when tiling is applied to the IFMap, as is commonly done when the entire IFMap size is too large for the SRAM, only a tile of IFMap is stored in the SRAM at a time \textit{using this layout}.

With such an SRAM layout, let us explain how to lower a $3 \times 3$ \conv{} layer to a GEMM. In the first 9 cycles, \textit{columns} of \texttt{1A, 1B, 1C, 2A, 2B, 2C, 3A, 3B, 3C} are read out from the SRAM, one column (word) per cycle. These 9 columns correspond to the first sliding window in the IFMap that the filter operates on (assuming no padding). In the next 9 cycles, columns corresponding to the next sliding window (with a stride of 1), i.e., \texttt{1B, 1C, 1D, 2B, 2C, 2D, 3B, 3C, 3D}, are read out from the SRAM and fed to the GEMM engine, one column at a cycle. This process repeats until all the sliding windows in the current IFMap tile resident in the SRAM finish, at which point the next tile starts until all the IFMap tiles finish.

Critically, while each IFMap element is read multiple times, it is deterministically sent to one fixed PE throughout the execution. For instance, all the 1C elements are read three times, but they are always read as a whole word and sent to the corresponding PEs each time. In this way, the entire on-chip SRAM can be organized as a single bank with a word size of 8 elements (e.g., 8 Bytes if INT8 is used). 
Note that this requires the channel size to be a multiple of word size, the implication of which is discussed in \Sec{sec:tpu}.

\begin{figure}[t]
    \vspace{-0.1cm}
    \centering
    \includegraphics[width=\linewidth]{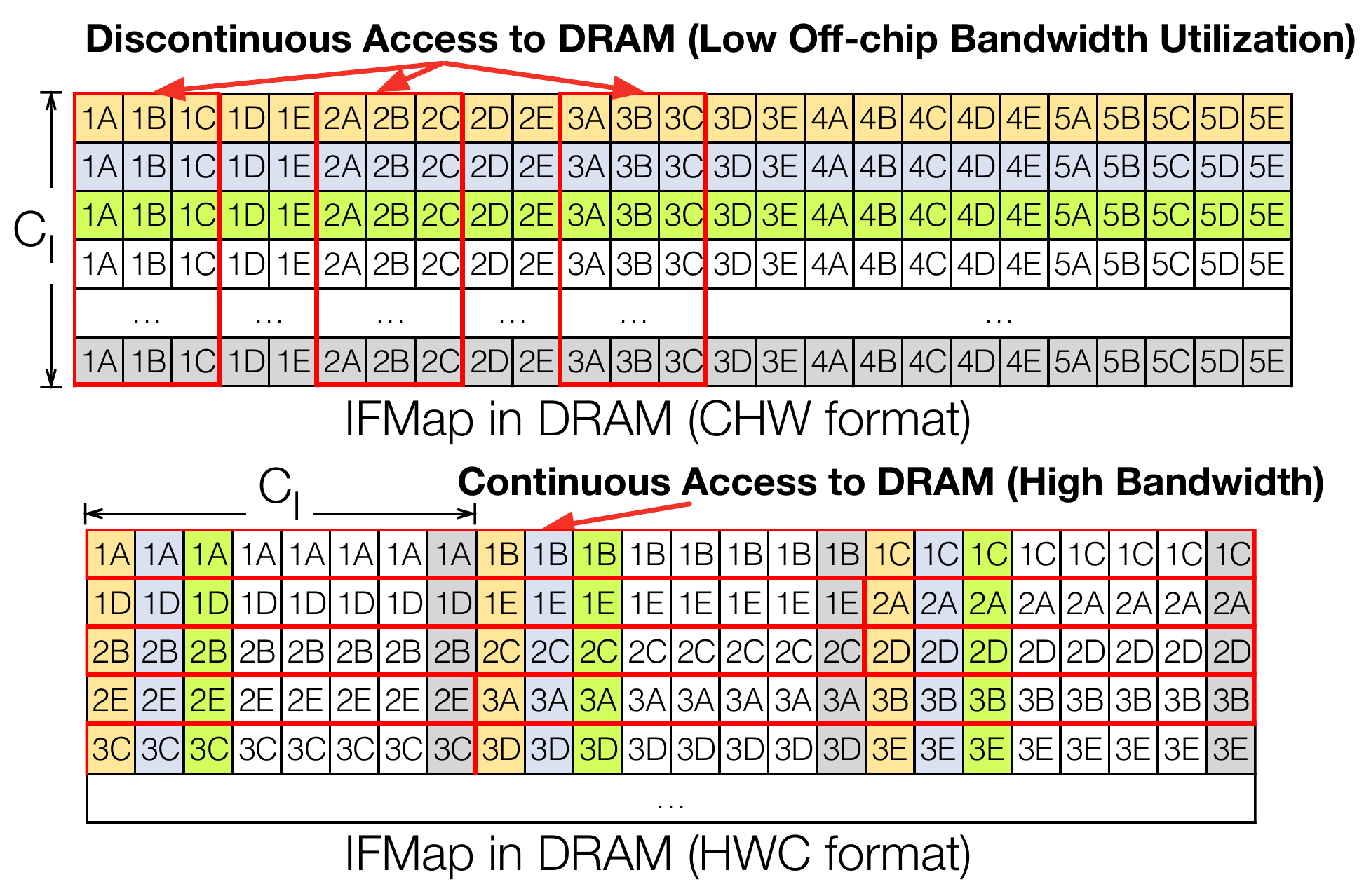}
    \vspace{-0.6cm}
    \caption{\small Advantage of $HWC$ format over commonly used $CHW$.}
    \label{fig:c_first_im2col_dram_example}
    \vspace{-0.5cm}
\end{figure}

\paragraph{DRAM Layout}
To maximize the DRAM bandwidth utilization when loading the IFMap data to the SRAM with the HWC layout, we propose to store the IFMap in the DRAM using the HWC format instead of the commonly used CHW format.
\Fig{fig:c_first_im2col_dram_example} compares the access patterns to produce the lowered matrix tile example in \Fig{fig:benefit_c_first_im2col} between the two DRAM layouts.

The access pattern to the CHW-based IFMap contains discontinuous accesses (e.g., \texttt{1A-1C}, \texttt{2A-2C}, \texttt{3A-3C} in the first channel) to the DRAM, which under-utilizes the off-chip memory bandwidth and increases the SRAM filling latency.
In contrast, the access pattern to the HWC-based IFMap contains mostly continuous accesses (e.g., eight channels of \texttt{1A} to \texttt{1C}), which better utilizes the off-chip bandwidth.

Note that when stride=1 the gap between the CHW and HWC format is small because the W-dimension (e.g., 128) is typically large enough for using the DRAM bandwidth.
However, the HWC format is critical to resolve the performance issue of larger strides, as we explain later.

\paragraph{General Principle} 
Inherently, the above approach simply changes the way lowered IFMap elements are arranged in existing \im{}. \Fig{fig:im2col_hw_vs_c} illustrates the difference between the channel-first method and existing channel-last method. Note that this reordering is conceptual, as the lowered IFMap never physically exists --- it is dynamically generated and consumed.

Recall from \Fig{fig:im2col} that in \im{} the IFMap is converted to a [$H_O W_O \times H_F W_F C_I$] lowered IFMap. In the existing channel-last \im{} approach, the $H_F W_F C_I$ dimension of the lowered IFMap is expanded in the $C_I \rightarrow H_F \rightarrow W_F$ order, which stores the $H_F\times W_F$ elements in a sliding window across all channels sequentially. In contrast, our approach constructs the $H_F W_F C_I$ dimension in the $ H_F \rightarrow W_F \rightarrow C_I$ order, which stores elements of the same position across the $C_I$ channels sequentially. Intuitively, our lowered IFMap simply shuffles the columns of the lowed IFMap in the existing method.

In this sense, the correctness of the channel-first method can be understood as: changing the column order in a matrix does not change the result of GEMM (so long as the other matrix elements are reordered accordingly).

\begin{figure}[t]
    \centering
    \includegraphics[width=\linewidth]{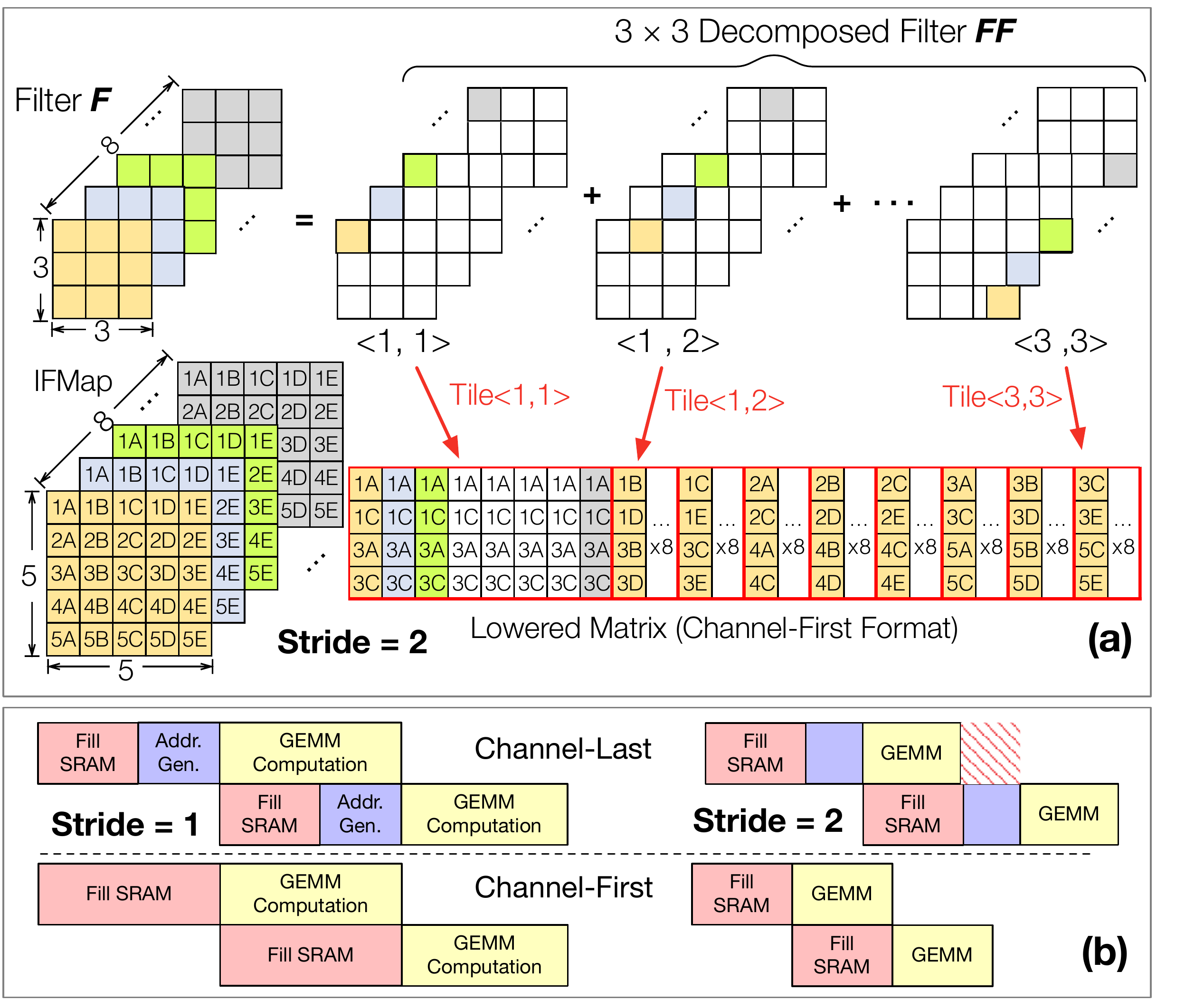}
    \vspace{-0.2cm}
    \caption{\small Example for supporting \conv with a stride of two.
   (a) We tile the lowered matrix according to the filter decomposition, where each tile corresponds to a $1 \times 1$ \conv{} layer. 
   (b) Our channel-first \im{} is insensitive to stride because it requires simple address generation and the on-chip SRAM filling time also reduces as the stride increases from 1 to 2.
   }
    \label{fig:c_first_im2col_memory_example}
    \vspace{-0.4cm}
\end{figure}

\subsection{Supporting Conv Variants}
\label{subsec:conv_variant}

We will first describe a particularly useful way of understanding the proposed channel-first \im{}, using which we then explain how this algorithm can be trivially extended to support \conv{} variants, such as the strided convolution, which are inefficient using existing implicit \im{} method.

\paragraph{Decomposed $1\times 1$ CONVs} The studied implicit \im{} method essentially decomposes the $H_F\times W_F\times C_I$ filter into $H_F\times W_F\times C_I$ $1\times 1$ filters (and properly accumulating the partial sums), because each column of data that enters the GEMM engine consists of elements from the same position across all channels, effectively performing $1\times 1$ \conv{}s.

Critically, the $1\times 1$ \conv{}s can be performed in an arbitrary order due to the commutativity of accumulation. We propose to iterate over the decomposed filters (as opposed to over sliding windows). We first compute the $1\times 1$ \conv{}s associated with the first decomposed filter and then move to the $1\times 1$ \conv{}s associated with the next decomposed filter, and so on. Consider the example in \Fig{fig:c_first_im2col_memory_example}a. The first decomposed filter $\langle 1, 1 \rangle$ convolves with IFMap data at \texttt{1A, 1C, 3A}, and \texttt{3C}; the next decomposed filter, say, $\langle 1, 2 \rangle$ convolves with IFMap data at \texttt{1B, 1D, 3B}, and \texttt{3D}. The lower-right corner in \Fig{fig:c_first_im2col_memory_example}a shows the corresponding lowered IFMap.

The channel-first algorithm is naturally insensitive to the stride size, because the SRAM filling latency and the GEMM latency decrease simultaneously and proportionally for each tile.
\Fig{fig:c_first_im2col_memory_example}b illustrates the difference between the existing channel-last method and our channel-first method as the stride changes from 1 to 2.
The performance of the existing method degrades because the GEMM latency of a tile decreases but the SRAM filling latency remains the same.
In the channel-first method, however, when the stride decreases, the size of each lowered IFMap tile proportionally decreases, e.g., from $9\times 8$ (\Fig{fig:benefit_c_first_im2col}) to $4\times 8$ (\Fig{fig:c_first_im2col_memory_example}a).
Thus, the GEMM latency can still hide the SRAM filling latency, without hurting performance.

\section{Support for Systolic Array}
\label{sec:tpu}

In this section, we describe how to support the proposed implicit channel-first \im{} method on systolic array using Google's TPU architecture as the base design. We first describe the basic idea of mapping the channel-first \im{} algorithm to the TPU (\Sec{sec:tpu:algo}), followed by optimizations that further improve the mapping efficiency (\Sec{sec:tpu:opt}).

Our measurement results show that it's very likely that the TPU adopts the same algorithm and same set of optimizations.
To our best knowledge, we are the first to show how to support implicit \im{} in a systolic array in the public domain. That said, we will briefly discuss why we suspect that the TPU design is similar to what we describe here, and how the baseline design needs specific optimizations for the higher performance in certain scenarios.

\begin{figure}[t]
    \begin{center}
    \vspace{-0.2cm}
    \includegraphics[width=\linewidth]{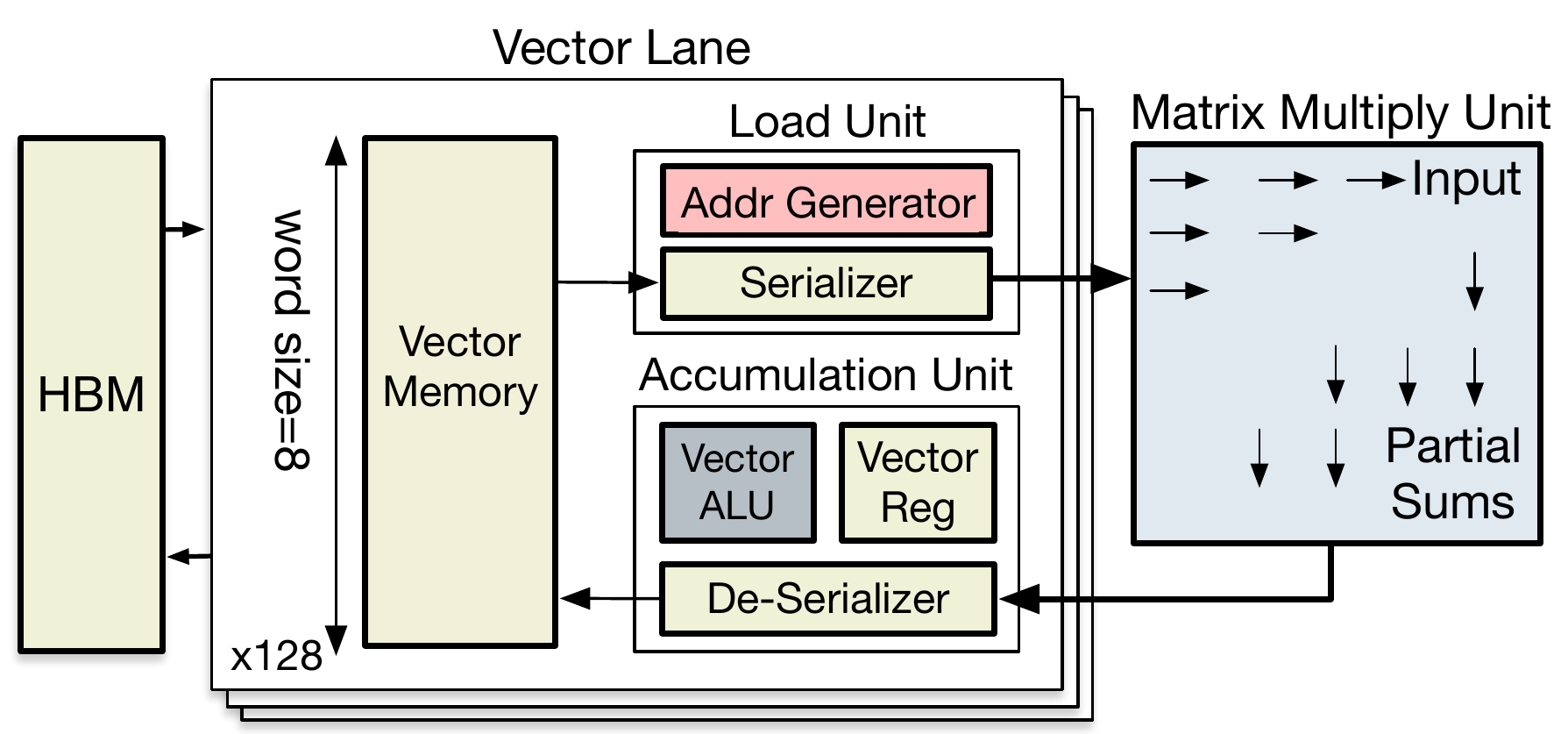}
    \end{center}
    \vspace{-0.2cm}
    \caption{\small The simplified TPU-v2 core architecture details~\cite{tpu_cacm,norrie21tpu}.}
    \vspace{-0.4cm}
    \label{fig:tpu_arch}
\end{figure}

\subsection{Basic Algorithm Mapping}
\label{sec:tpu:algo}

\paragraph{Baseline TPU}
We model the baseline systolic array architecture after the published details of Google's TPU-v2/v3 architecture~\cite{tpu_cacm,norrie21tpu}.
In particular, we consider a baseline dual-core TPU-v2. 
TPU-v3 is similar in the core microarchitecture but uses different core numbers.
\Fig{fig:tpu_arch} shows the details of a single TPU core, which contains a $128 \times 128$ weight stationary systolic array for GEMM. We describe the architectural details that are most relevant to implementing our \im{} algorithm.

Different from the separate buffer design in TPU-v1~\cite{tpu_paper}, TPU-v2 uses a unified on-chip SRAM for storing IFMap, weights, and OFMap. The unified SRAM is split into 128 different SRAM arrays, each dedicated to exchanging data with \textit{a fixed PE row in the systolic array}. That is, this is \textit{not} a 128-bank SRAM with a 128 $\times$ 128 cross-bar, but 128 separate SRAM arrays, each with a single read/write port.

The TPU-v2 chooses a word size of eight for each SRAM array, which is thus called a vector memory in \Fig{fig:tpu_arch}. For instance, if FP16 or BFloat16 is to be used, each access to a vector memory would be reading/writing 16 bytes.

\paragraph{Implementing Implicit im2col}
The key challenge of implementing the aforementioned channel-first implicit \im{} method on the TPU is that the TPU, as a systolic array, has a time-delayed data access pattern. One naive way to address it would be to skew the data layout in the SRAM and DRAM described in \Sec{sec:algo}. However, it would lead to frequent skewing and restoring for other non-GEMM layers such as pooling and batch normalization~\cite{batchnorm}.

Instead, the channel-first implicit \im{} can be trivially mapped to the TPU by leveraging the unique SRAM organization of the TPU, which uses 128 independent SRAM arrays, each for a PE row. We simply map each row of our proposed $HWC$ SRAM layout (\Fig{fig:c_first_im2col_memory_example}a) to an SRAM array. Naturally, each SRAM array stores a single channel in the IFMap. Since each SRAM array is independent, we would simply use an address generation logic to generate the appropriate address to each SRAM array such that the 128 addresses are skewed by one cycle to fit the systolic data flow. That is, instead of skewing the data layout, we skew the address generation, which is enabled by the SRAM organization in the TPU.

\paragraph{Leveraging Large Word Size}
Each of 128 SRAM arrays in the TPU has a word size of 8. Therefore, each time we read from an SRAM array, it would return 8 data elements (whether it is FP16 or BFloat16). To utilize the large word size, our idea is to put different inputs from the same batch into the same SRAM array. That is, each SRAM array would store a channel of the IFMap from 8 different inputs. We call this new data layout $HWCN$, where $N$ denotes the batch dimension.

A potential issue with the large word size is that, while an SRAM array can produce 8 data elements, each PE in the GEMM engine would accept only one data element per cycle. Therefore, we propose to use a buffer (the serializer in \Fig{fig:tpu_arch}) in-between the SRAM array and the GEMM engine to hold all the 8 data elements read at once; the serializer would then issue one element to the GEMM engine each cycle. Accordingly, we would read data from each SRAM array and update the serializer only once every 8 cycles, reducing the SRAM switching activity.

\begin{figure}[t]
    \begin{center}
    \includegraphics[width=\linewidth]{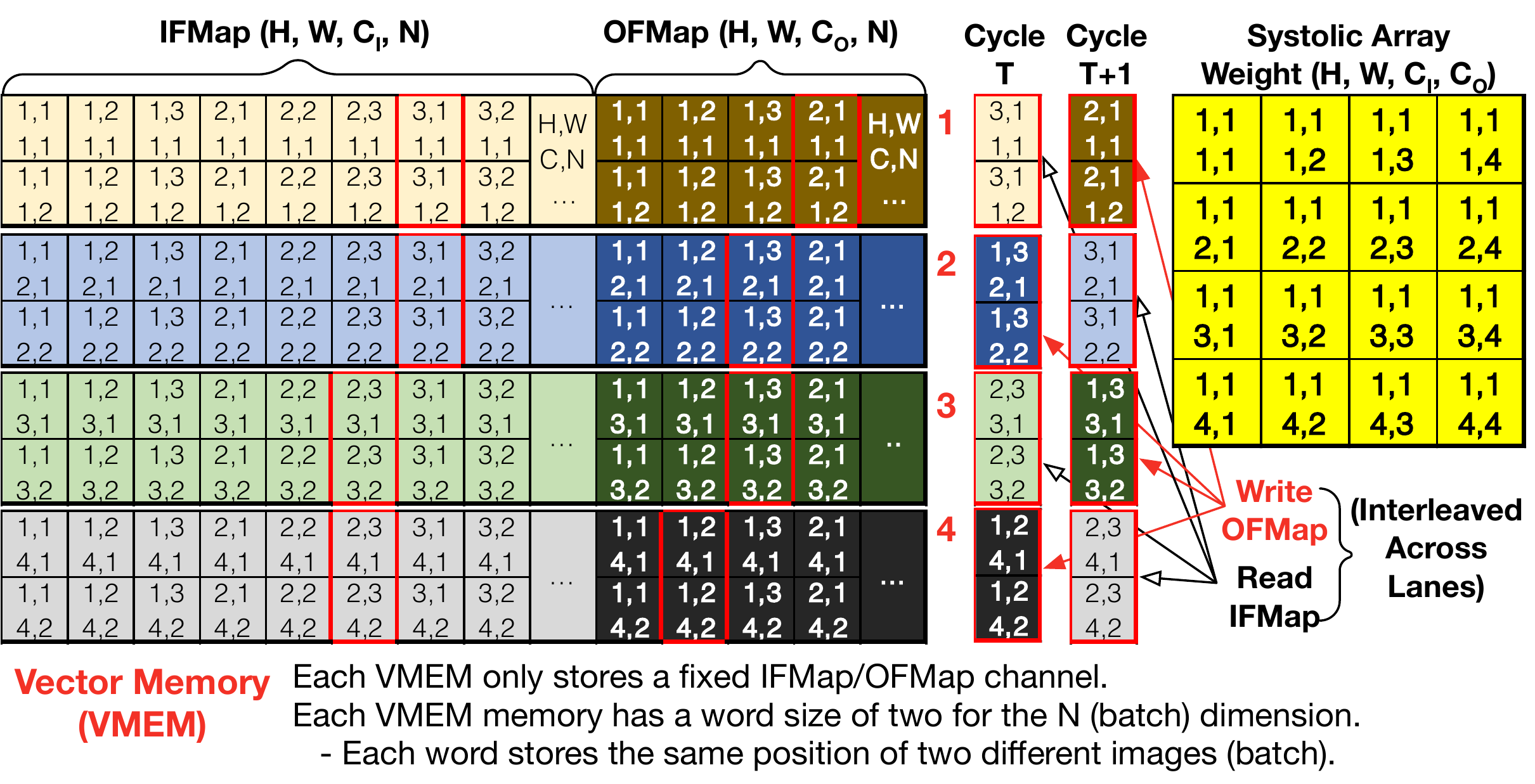}
    \end{center}
    \vspace{-0.2cm}
    \caption{\small The example of executing the tile $\langle 1,1\rangle$  for IFMap $N=2, C_I = 4, H_I=C_I=5$, filter $H_F=W_F=3$, OFMap ($H_O=C_O=3$) on the $4\times 4$ weight stationary systolic array with 4 separate SRAMs. Each SRAM has a word size of 2, and stores both IFMap and OFMap.}
    \label{fig:tpu_example}
    \vspace*{-0.5cm}
\end{figure}

\paragraph{Leveraging the Unified Memory} The on-chip memory in the TPU is unified in that it stores both IFMap and OFMap. As a result, storing to the OFMap and reading from the IFMap will contend the same port of each SRAM array. 
We address this issue by leveraging the fact that each SRAM array is read only once per 8 cycles, which allows us to interleave storing the OFMap data to an SRAM array with the load, effectively posing zero contention/overhead.
For implementation, we add a de-serializer for each vector memory in \Fig{fig:tpu_example}, which receives the result from the systolic array every cycle and writes to the vector memory every 8 cycles.

\paragraph{Example} 
Without loss of generality, we use a small working example in \Fig{fig:tpu_example} to illustrate the on-chip vector memory layout and its interaction with the systolic array when executing the tile $langle 1,1 \rangle$ for the filer size of $3\times 3$.
The indices in a vector memory element indicate the height, width, channel, and batch, correspondingly.
Each column in a vector memory is a word, e.g., $(3,1,1,1)$ and $(3,1,1,2)$ in the first vector memory.
Because we are currently executing the GEMM for tile $\langle 1,1\rangle$ , we only need to store each $\langle 1,1\rangle$  position of the $C_I \times C_O$ (i.e., $4 \times 4$) filters in the systolic array.

Because this example has a word size of two, each vector memory is read for IFMap or written for OFMap every two cycles in an interleaved fashion.
For example, at the cycle $T$, the first and third row of the systolic array reads a word from the corresponding vector memory, while the second and fourth column (the partial sum results comes from the bottom PEs in \Fig{fig:tpu_arch}) writes to the vector memory.
At cycle $T+1$, each vector memory switches to read or write.

\begin{figure}[t]
    \begin{center}
    \includegraphics[width=\linewidth]{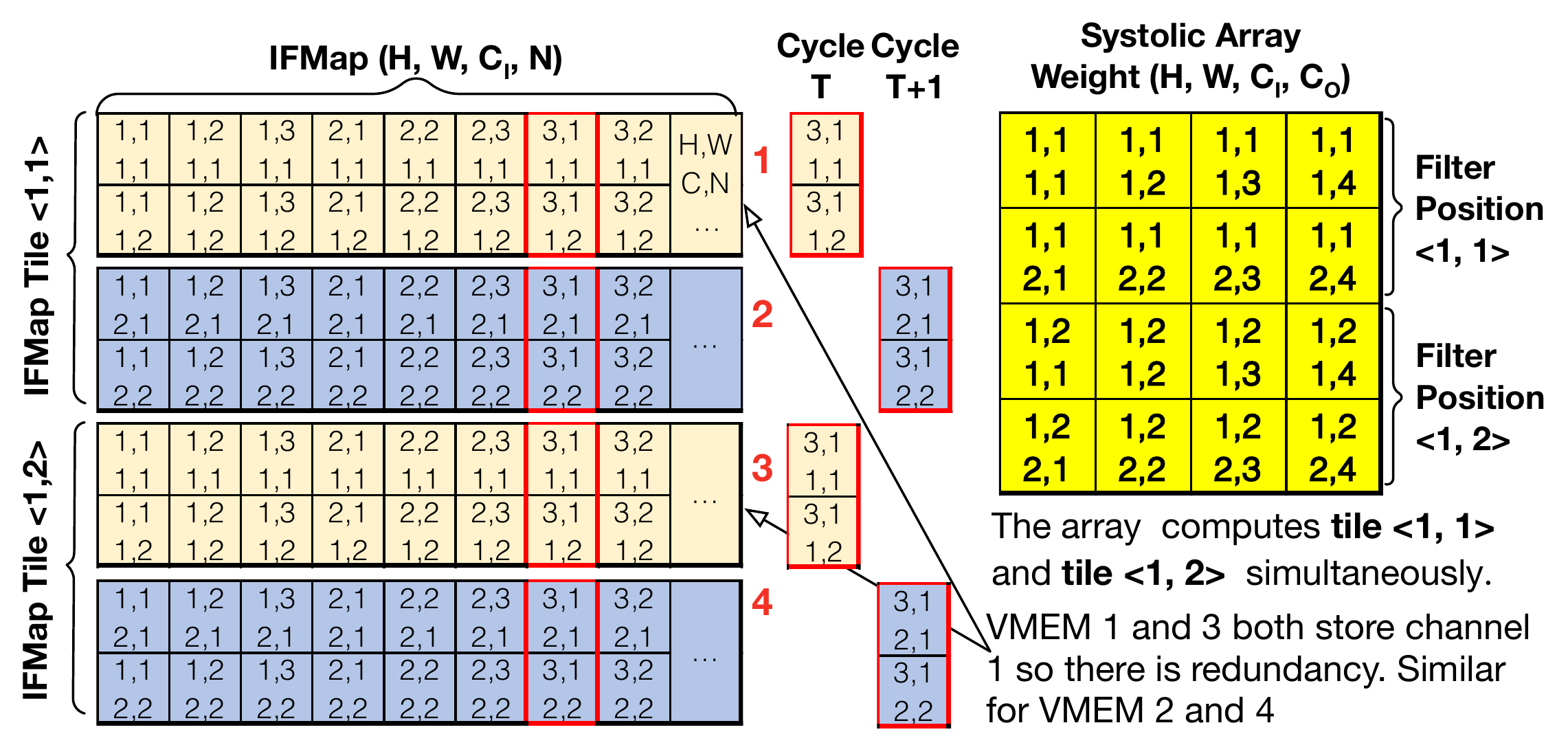}
    \end{center}
    \vspace{-0.3cm}
    \caption{\small The example of multi-tile computation (tile $\langle 1,1\rangle$  and $\langle 1,2\rangle$ ) for \conv{} with $N=2, C_I = 2, H_I=C_I=5, H_F=W_F=3, H_O=C_O=3$, which doubles the array utilization compared to the single-tile computation. It leads to input duplication (e.g., $2 \times$) in the SRAM.}
    \vspace{-0.4cm}
    \label{fig:cin_tpu}
\end{figure}

\subsection{Optimizations}
\label{sec:tpu:opt}

\paragraph{Limitations}
The design described above becomes inefficient when the input channel size is small, e.g., 3 in the first layer of today's CNNs. This is because each PE row in the systolic array reads from one SRAM array, which stores one input channel in the IFMap. An IFMap channel of 3 leaves 125 PE rows idle, i.e., leading to a severe array under-utilization.

As a result, the design described before requires the channel size to be padded to 128, respectively. We introduce one nifty optimization that avoids these wastes.

\paragraph{Multi-tile Computation}
To mitigate array under-utilization and hence performance loss when the input channel is small, we propose to fill the vector memories with data from other tiles (yet to be computed), essentially concurrently computing multiple tiles. For instance, the original method in \Fig{fig:tpu_example} only computes a single tile $\langle 1,1\rangle$  out of the total $H_F \times W_F$ tiles.
When the channel size is small, say half of the array size, we can compute the tile $\langle 1,1\rangle$  and $\langle 1,2\rangle$  simultaneously.

\Fig{fig:cin_tpu} shows an example when IFMap channel size $C_I$ is 2 and the array size is 4.
The systolic array stores all the weight elements from the $\langle 1,1\rangle$  and $\langle 1,2\rangle$  positions of $2 \times 4$ (i.e., $C_I \times C_O$) filters, and therefore computes the feature matrix tile $\langle 1,1\rangle$  and $\langle 1,2\rangle$  at the same time.
This optimization is essentially merging the two tiles to form a larger tile so the correctness is guaranteed by the associativity of GEMM. 

This optimization does not require any hardware modification as its only difference from the single-tile computation is the vector memory filling and address generation, which can be supported via instructions.
Meanwhile, the tile data replication is performed before IFMaps flow into the systolic array, so no additional synchronization is needed.

However, this optimization also leads to the IFMap duplication in the on-chip vector memory. 
For example, computing tile $\langle 1,1\rangle$ and tile $\langle 1,2\rangle$ at the same time requires storing the same channel twice inside the vector memory in \Fig{fig:cin_tpu}.
The maximum number of tiles allowed is a parameter for the tradeoff between memory overhead and performance improvement, which we explore in the evaluation section.

\subsection{Discussions}
\label{sec:tpu:disc}

Our educated guess is that the design described in \Sec{sec:tpu:algo} is similar to what is implemented in the TPU. This will be evident in our evaluation.
We suspect that the reason the TPU uses separate SRAM arrays is exactly to allow each SRAM array to be individually addressed so as to feed data to the GEMM engine in a time-delayed fashion.

The reason for each SRAM array to have a large word size is to amortize the area/power overhead. It is established that a small SRAM with a narrow word is inefficient. 
As we will show in \Sec{sec:eval}, for an SRAM array of size 256~KB, having a word size of 4 bytes increases the area overhead by 3.2 times compared to that when the word size is 32 bytes.
TPU design is clever in leveraging the large word size through batching, which is common in training --- a key focus of TPU-v2/v3.

\section{Implementation on Tensor Cores}
\label{sec:gpu}

\begin{figure}[t]
    \begin{center}
    \includegraphics[width=.95\linewidth]{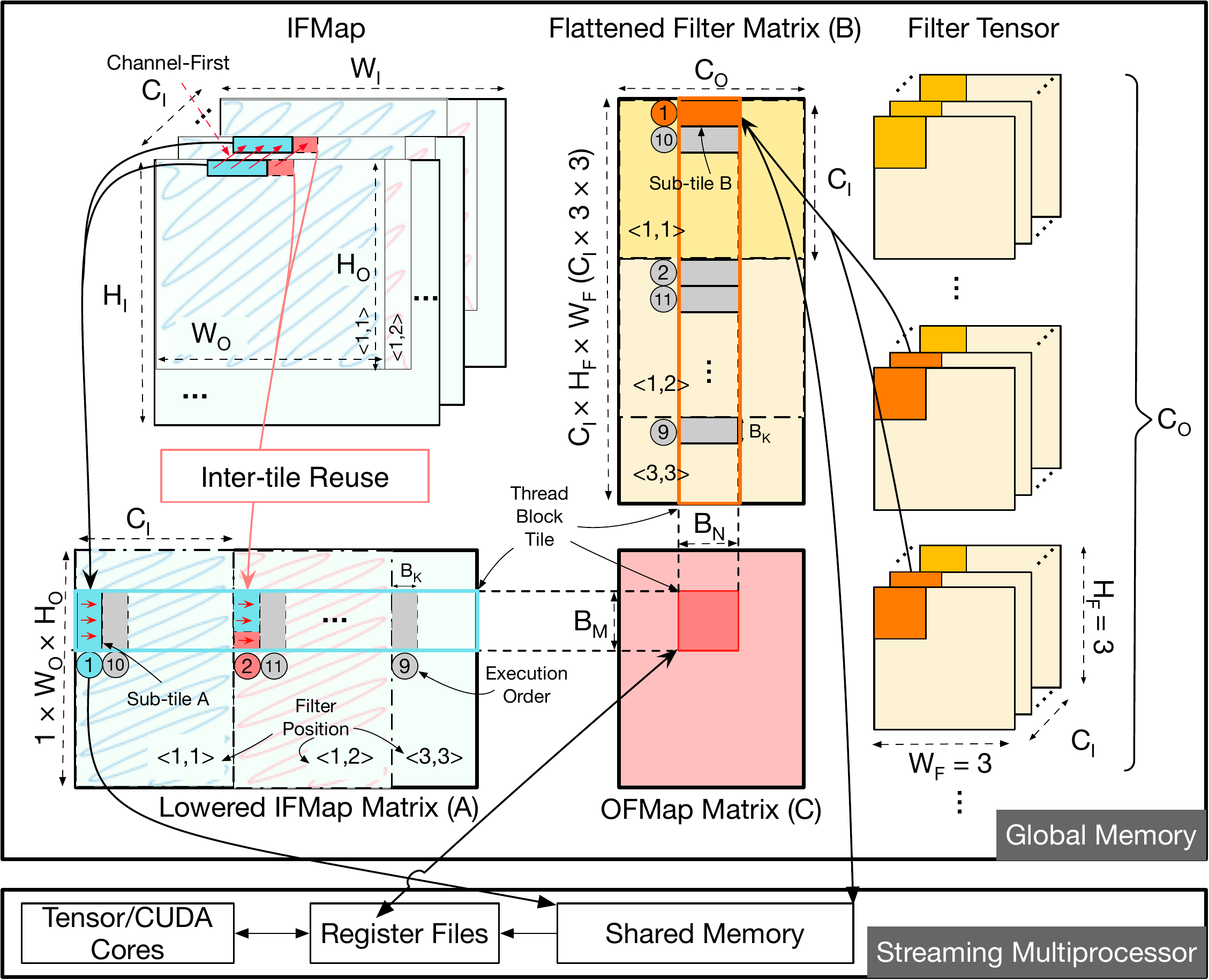}
    \end{center}
    \vspace{-0.2cm}
    \caption{\small Block-level implicit channel-first \im{} on GPU TCs.}
    \label{fig:gpu_implicit}
        \vspace{-0.3cm}
\end{figure}

In this section, we describe how to implement the channel-first implicit \im{} on GPU with Tensor Cores. The TCs~\cite{v100,raihan2019modeling} adopt dot-product units for GEMM acceleration, which exhibit regular access pattern instead of the TPU's time-delayed access pattern. Therefore, the channel-first implicit \im{} is naturally amenable to GPU implementation.

\paragraph{Block-level Channel-First im2col}
The challenge in applying our method to the TC is that there are usually many TCs on a GPU (eight TCs per streaming multi-processors/SM), so it is necessary to parallelize the GEMM to achieve high-performance on the GPU.
To avoid the atomic update to the output matrix, GPU typically partitions the output matrix and assigns an output matrix tile to a thread block (TB). Thus, different TBs run on different SMs in parallel.

However, this parallelization approach is incompatible with the straightforward version of the channel-first implicit \im{} algorithm.
The reason is that the channel-first algorithm relies on filer decomposition, which needs to accumulate the OFMap $H_F \times W_F$ times and thus requires atomic updates.

We address the above challenge by simply applying our implicit \im{} method at the blocked GEMM-level.
\Fig{fig:gpu_implicit} illustrates the basic idea, which is to perform the channel-first \im{} after the equivalent GEMM is blocked. 
In this way, different thread blocks still update different parts of the OFMap matrix, which avoids the expensive atomic update.

\paragraph{Inter-tile Reuse}
GPU's on-chip SRAM per SM is small compared to that of the TPU, and competition for resources between thread blocks assigned to the same SM could degrade the thread-level parallelism~\cite{kim2020duplo}.
It is thus equally important to increase the utilization of the SRAM to avoid the blocked GEMM being bounded by the off-chip DRAM bandwidth.

We propose a reordering technique to increase data use and thus SRAM utilization.
Our critical observation is that when the filter size is smaller than the stride size (which is true to virtually all CNNs today), the corresponding tiles of different decomposed filters have significant overlaps. For instance in \Fig{fig:c_first_im2col_memory_example}c, half of the IFMap data required by decomposed filter $\langle 1, 1\rangle$ and $\langle 1, 3\rangle$ overlap (i.e., elements at 1C and 3C). When the IFMap size increases to $99\times 99$, the working set overlap between these two decomposed filters becomes 96\%.

This overlap allows us to reduce the SRAM filling latency by reordering the decomposed filters.
A naive execution order of the above block-level \im{} is to iterate over decomposed filters as they show up on the original filter, which is equivalent to iterating over the lowered IFMap tile of the same filter position and then move to the other position.
For the example in \Fig{fig:gpu_implicit}, this would fetch the subtile of matrix A with the order of \circledwhite{1},\circledwhite{\small 10}, ...\circledwhite{2}, \circledwhite{\small 11}, which has no data reuse.
Instead, a simple reordering to \circledwhite{1}, \circledwhite{2}...\circledwhite{\small 10},\circledwhite{\small 11}... would exploit the data reuse between subtiles. We leave it to future work to design an optimal reordering strategy.

\begin{table}[t]
    \centering
    \caption{TPU-v2 simulator configurations.}
    \small
    \begin{tabular}{c| c}
      \hline
      \specialcell{\textbf{Compute}\\ \textbf{Unit}} & \specialcell{$128 \times 128$ Systolic Array @ 700Mhz\\ 256 Vector ALUs for partial sum accumulating} \\
      \hline
      \specialcell{\textbf{Regs}} & 256 Vector Regs,  $8 \times 4$ bytes for each reg \\
      \hline
      \specialcell{\textbf{On-chip}\\ \textbf{Memory}} & \specialcell{32~MB Unified On-chip Memory\\128 SRAMs with $8 \times 4$ bytes Word Size}\\
      \hline
      \specialcell{\textbf{Off-chip}\\\textbf{Memory}} & 700~GB/s High Bandwidth Memory \\
      \hline
    \end{tabular}
    \label{tab:archconf}
    \vspace*{-0.3cm}
  \end{table}
  
\section{Methodology}
\label{sec:methodology}

\begin{figure}[t]
    \vspace{-0.1cm}
    \centering
    \subfloat[\small GEMM. Average error: 4.42\%.]
    {
      \includegraphics[trim=0cm 0 0 0, clip, width=0.49\linewidth]{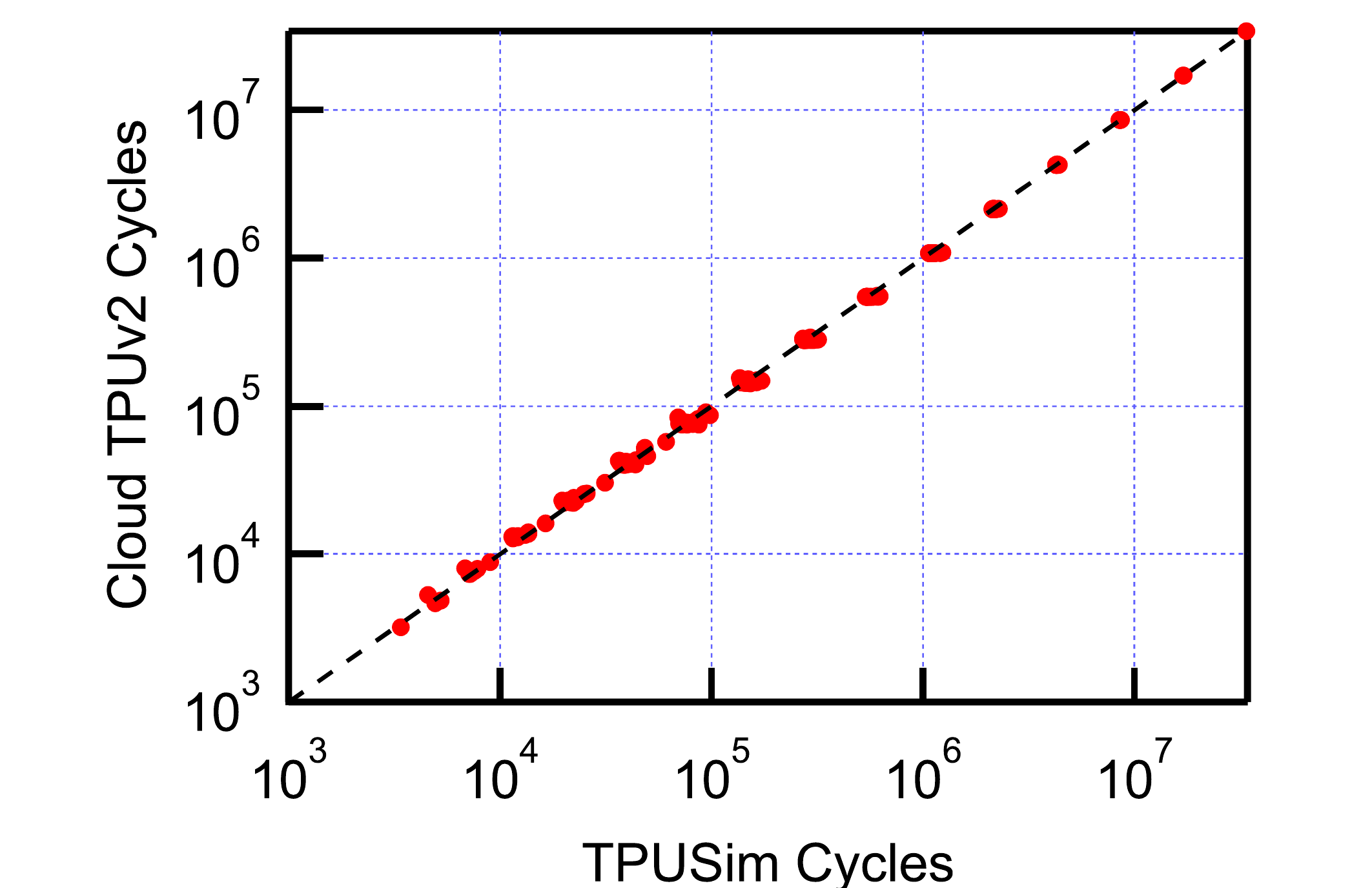}
      \label{subfig:gemm_validation}
    }%
    \subfloat[\small Conv. Average error: 4.87\%.]
    {
      \includegraphics[trim=0cm 0 0 0, clip, width=0.49\linewidth]{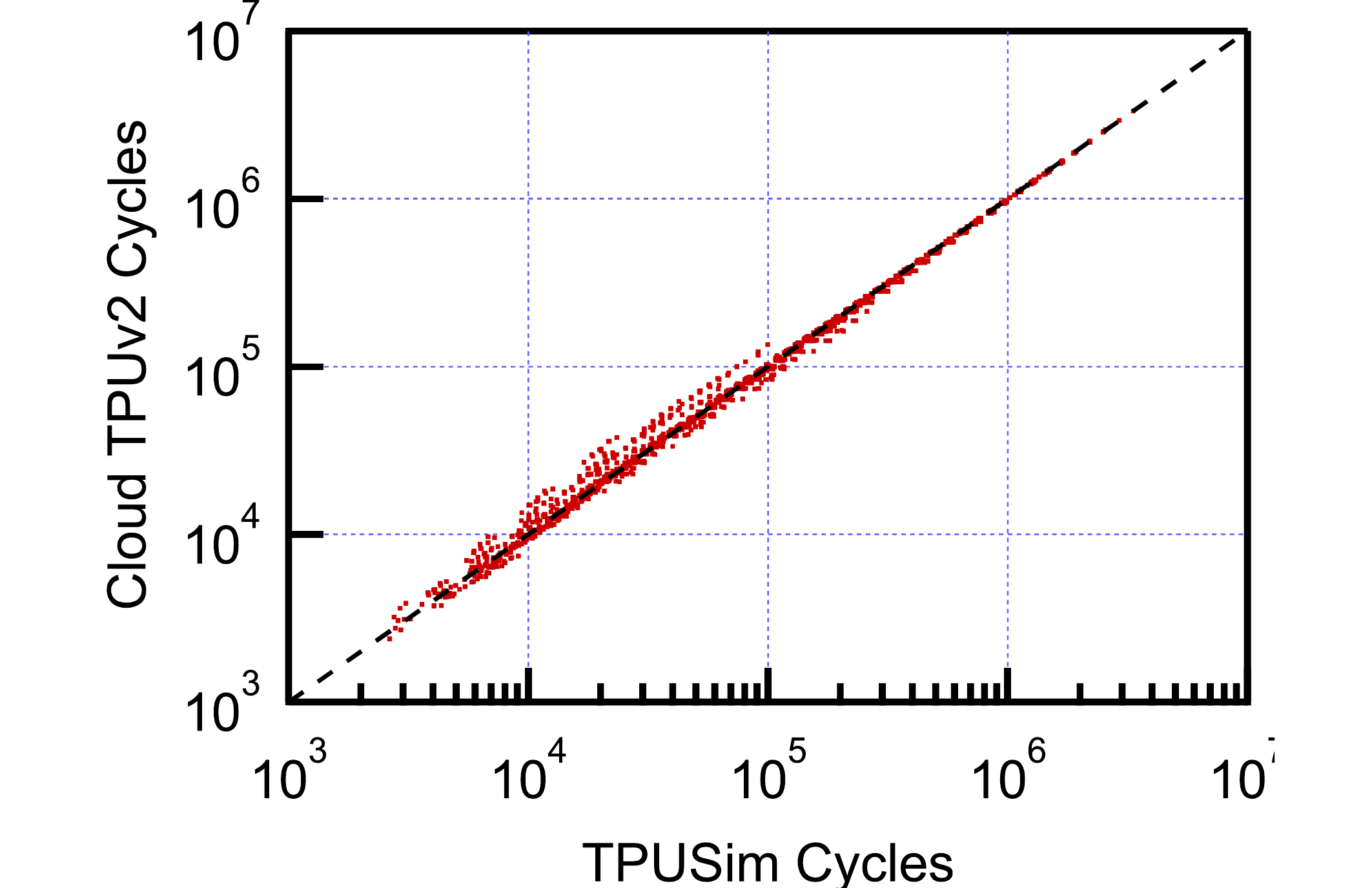}
      \label{subfig:conv_validation}
    }
    \vspace{-5pt}
    \caption{\small TPUSim and TPUv2 comparison on microbenchmarks.}
    \label{fig:tpusim_validation}
   	\vspace{-0.3cm}
\end{figure}

\paragraph{TPU Experiment Setup}
For the systolic array experiment, we design and implement a configurable cycle-accurate simulator TPUSim. We configured this simulator with the same parameters as TPUv2~\cite{norrie21tpu} shown in \Tbl{tab:archconf}. For the off-chip HBM and on-chip SRAM, we use DRAMSim3\cite{wang2005dramsim} and CACTI\cite{thoziyoor2008cacti} to obtain the access latencies, respectively.

\paragraph{GPU Experiment Setup}
For the GPU experiment, our evaluations are done on the NVIDIA Volta 100 GPUs using the FP16 data type. The software stack is CUDA 10.2, and the comparison baseline is cuDNN 7.
Our blocked GEMM baseline is implemented based on the \texttt{cudaTensorCoreGemm} kernel from NVIDIA CUDA SDK 11.3~\cite{guide2013cuda}.

\paragraph{Workload}
We evaluate 7 popular neural networks, AlexNet~\cite{krizhevsky2012imagenet}, DenseNet~\cite{huang2017densely}, GoogleNet~\cite{szegedy2015going}, ResNet~\cite{he2016deep}, VGG~\cite{simonyan2014very}, YOLO~\cite{redmon2016you}, and ZFNet~\cite{zeiler2014visualizing}, which cover tasks in different domains and models of different sizes. We report inference process with the widely used ImageNet dataset.

\paragraph{Baselines}
To verify the performance of our algorithm on the systolic array, we compared the experimental results of our simulator with TPUv2 with equivalent hardware parameters.
In the GPU experiments, we called the CUDNN\_CON-VOLUTION\_FWD\_ALGO\_IMPLICIT\_PRECOMP\_GEMM interface to test the performance of different cases running on Tensor Cores as our baseline for comparison.

\section{Evaluation}
\label{sec:eval}

\begin{figure}[t]
    \hspace{-0.4cm}
    \subfloat[\small Multi-tile parameter effect.]
    {
      \includegraphics[trim=0 0 0 0, clip, width=0.49\linewidth]{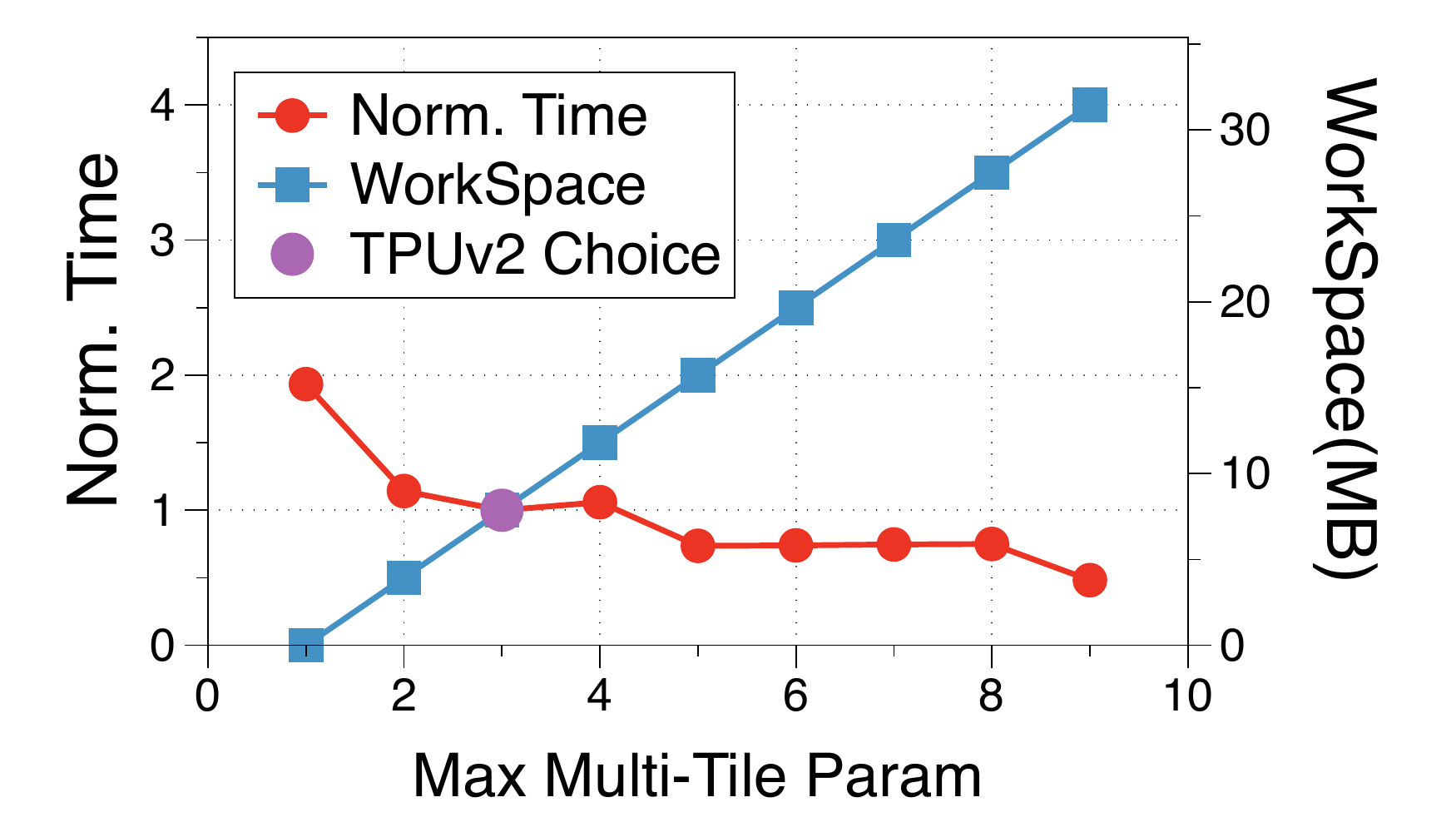}
      \label{subfig:sim_copy}
    }\hfill
    \subfloat[\small Multi-tile parameter validation.]
    {
      \includegraphics[trim=0 0 0 0, clip, width=0.49\linewidth]{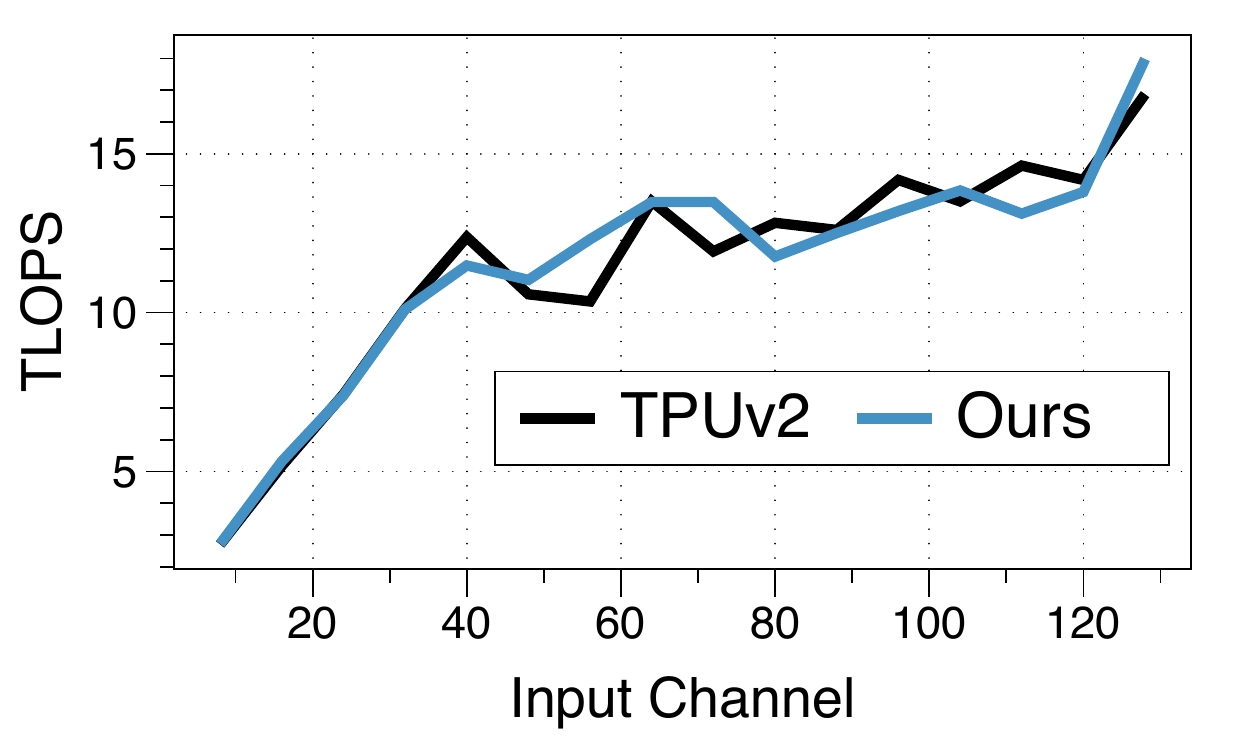}
      \label{subfig:tpu_sim_channel}
  
    }%
    \vspace{-3pt}
    \caption{\small The effect and validation of multi-tile parameter. (a) We use a \conv{} layer with $N=8, C_I=8, W_I=C_O=128, W_F=3$ and vary the number of tiles. (b) Validation of TPU's strategy (number of tiles equals $MIN(128/C_I, W_f)$) with an average error of 5.3\%.}
    \label{fig:TPU_Cin}
    \vspace{-0.3cm}
\end{figure}

\subsection{Evaluation on TPU}
\label{sec:eval:tpu}

\paragraph{TPUv2 Validation}
We validate the simulation accuracy of our simulator TPUSim against the TPUv2 using synthetic microbenchmarks.
We first perform validation for the GEMM primitive that TPU-v2 targets.
We vary the three parameters (e.g., $M, N, K$) of GEMM from 256 to 8192, and compare the execution cycles in the cloud TPU-v2 and TPUSim.
\Fig{subfig:gemm_validation} shows the validation results, where the averaged error for our simulation cycles for the GEMM primitive is 4.42\%.

We compare TPUSim with our implicit \im{} against TPUv2 using \conv{} layers that do not trigger our optimizations in \Sec{sec:tpu:opt}.
\Fig{subfig:conv_validation} shows the simulation cycle comparison with an averaged error of 4.87\%.
The close performance confirms our previous hypothesis that the TPU implements a similar implicit method.

\paragraph{Multi-Tile Parameter}
We first use a set of experiments to study the effectiveness and specific strategies of the multi-tile optimization (\Sec{sec:tpu:opt}) for input channel size less than 128.

We use a \conv{} layer with $N=8, C_I=8, W_I=C_O=128, W_F=3$ and vary the number of tiles in the multi-tile computation optimization.
\Fig{subfig:sim_copy} shows that the required on-chip vector memory workspace increases linearly as the maximum multi-tile param increases, but the performance improvement shows a diminishing return.
When the number of tiles is 3, our simulation results match with TPUv2.

We then repeat the above experiments with different channel and filter sizes.
We find that the TPU sets the number of tiles to $MIN(128/C_I, W_f)$: the multi-tile size is first bounded by the filter size and is just enough to occupy the $128\times 128$ systolic array.
Based on that strategy, \Fig{subfig:tpu_sim_channel} compares the TPUSim and TPUv2 performance in TFLOPS for varying input channel size, which has an average error of 5.3\%.
As such, we use this strategy in subsequent experiments.

\begin{figure}[t]
    \centering
    \subfloat[Model Result.]
    {
      \includegraphics[trim=0 0 0 0, clip, width=0.49\linewidth]{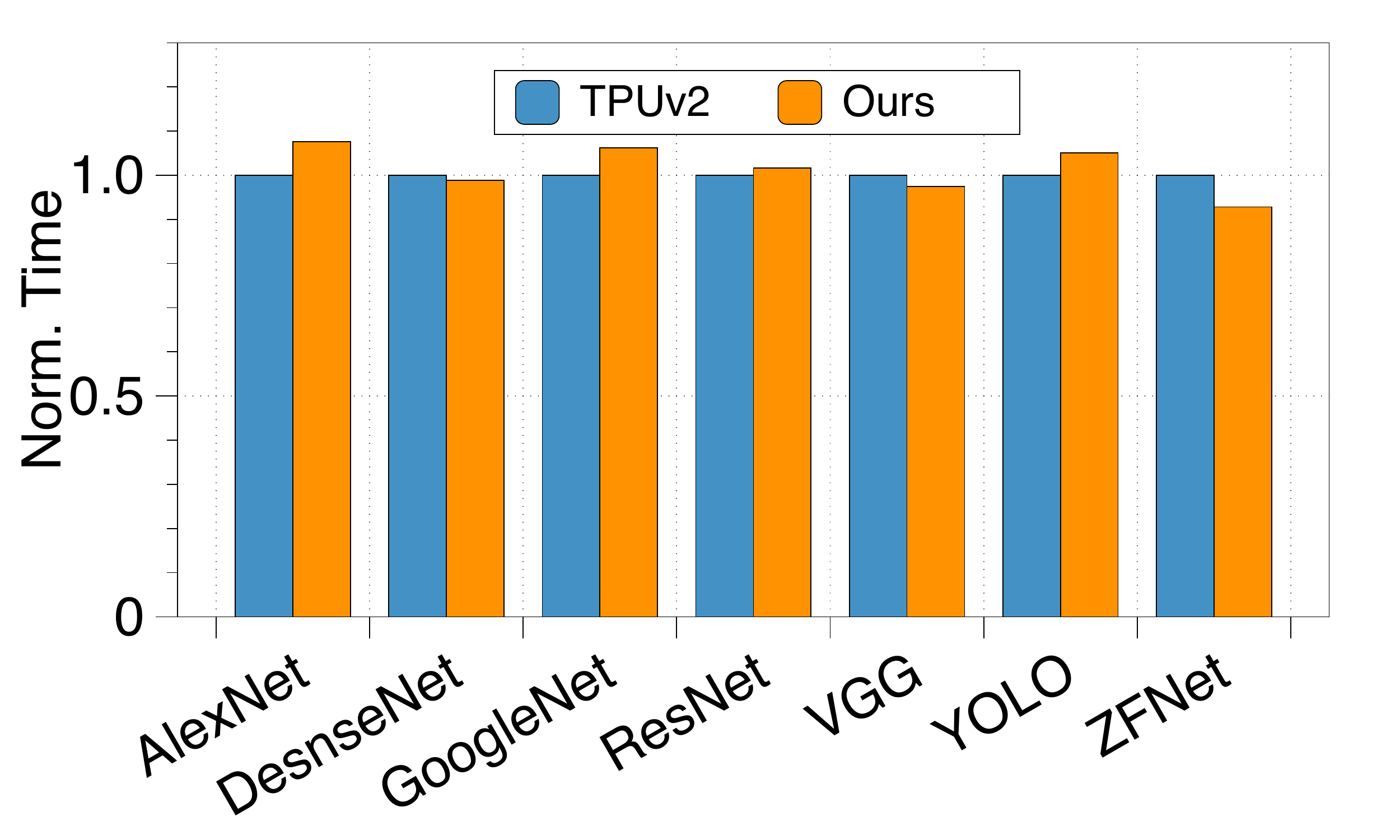}
    }
    \subfloat[Layer-wise error distribution.]
    {
      \includegraphics[trim=0 0 0 0, clip, width=0.49\linewidth]{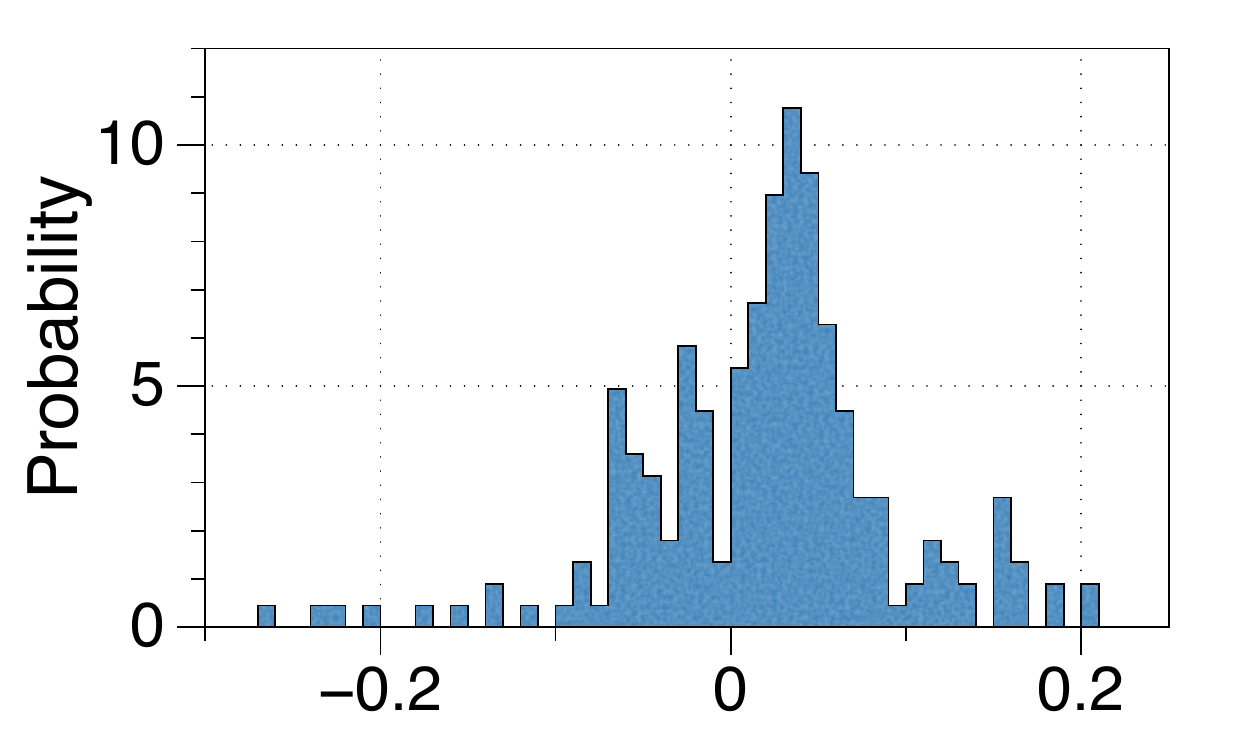}
    }
    \vspace{-5pt}
    \caption{\small Performance comparison between TPUSim and TPUv2 on DNN models at batch size of 8. The MAE for all layers is 5.8\%}
    \label{fig:TPU_time}
    \vspace{-0.5cm}
\end{figure}

\paragraph{End-to-end Model Results}
Besides synthetic \conv{} layers, we also compare the simulated and measured performance on real world CNN models that are described in \Sec{sec:methodology}. 
\Fig{fig:TPU_time}(a) shows that our design achieves performance results matched to TPUv2.
\Fig{fig:TPU_time}(b) shows the error distribution of simulated and measurement latency for all layers. 
The MAE of all layers is 5.8\%, which validates our algorithm and simulator.

\begin{figure}[t]
  \vspace{-0.1cm}
  \centering
  \subfloat[Systolic array size.]
  {
    \includegraphics[trim=.5cm 0 .5cm 0, clip, width=0.49\linewidth]{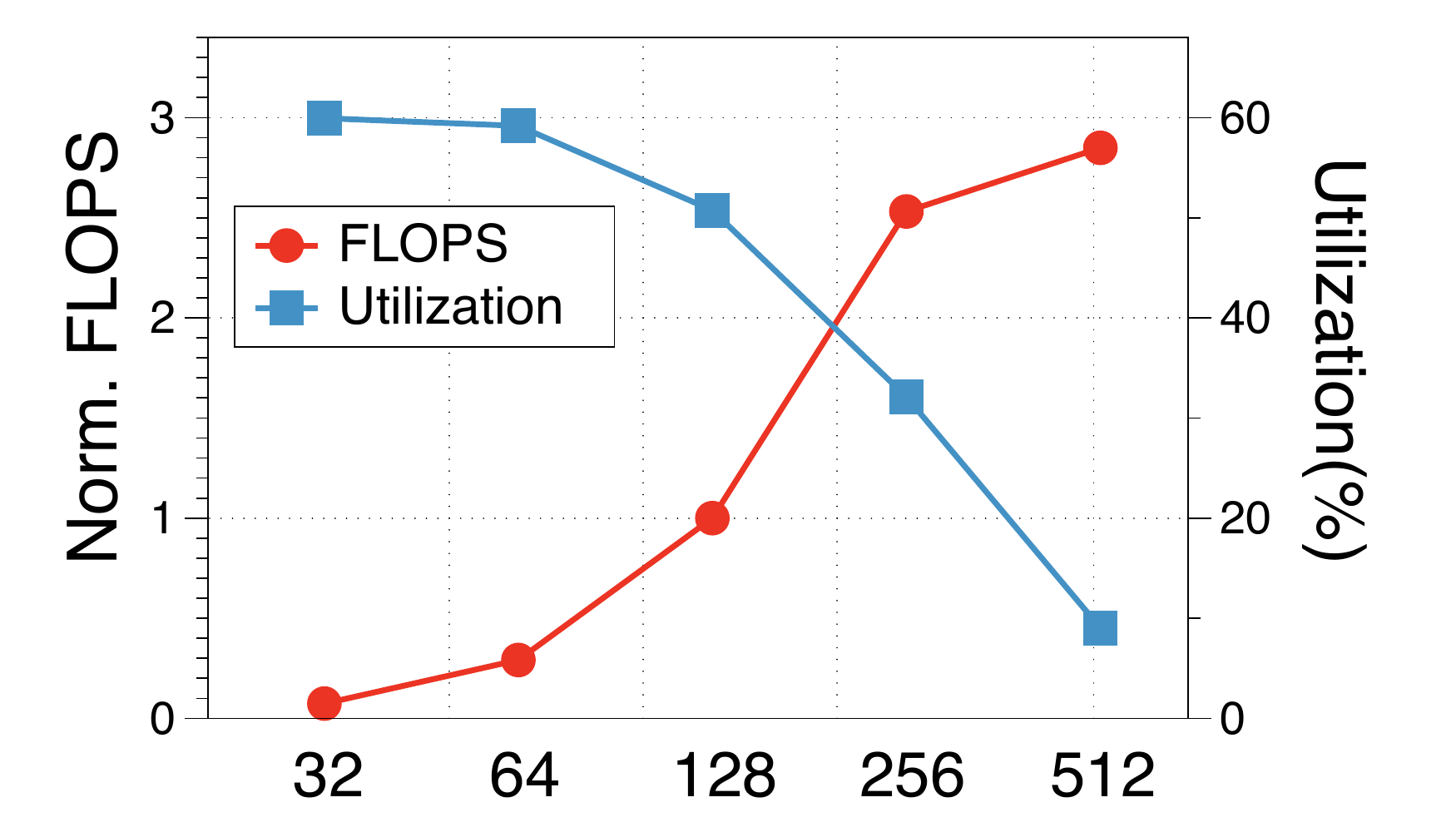}
    \label{subfig:eval_bank}
  }%
  \subfloat[Word size.]
  {
    \includegraphics[trim=.3cm 0 .5cm 0, clip, width=0.49\linewidth]{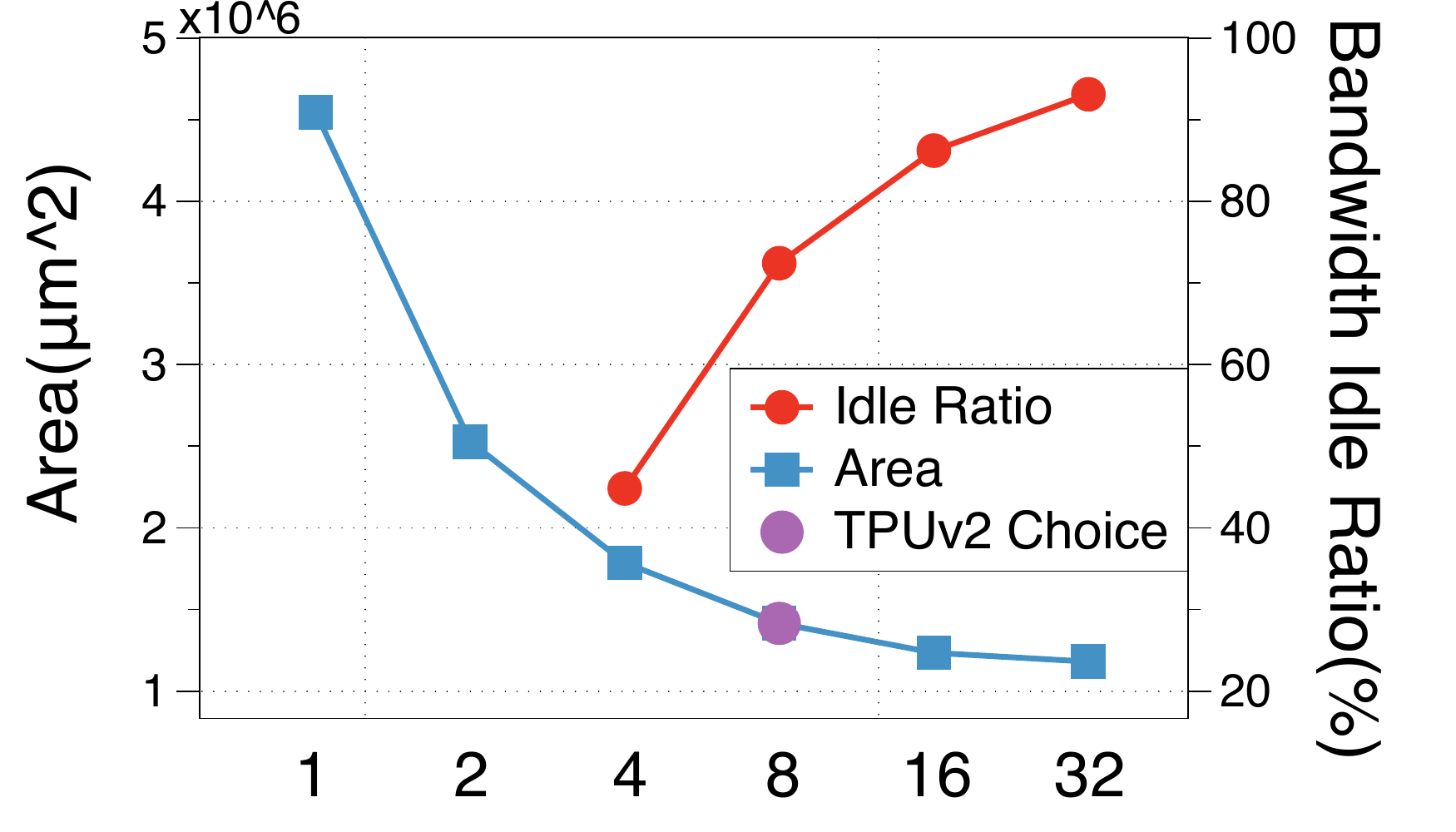}
    \label{fig:area}

  }
  \vspace*{-0.1cm}
  \caption{\small Hardware design space exploration using TPUSim.}
  \label{fig:eval_lane}
  \vspace{-0.3cm}
\end{figure}

\paragraph{Hardware Design Space}
With our validated simulator, we leverage its configurability to performing a design space exploration to understand the design decisions of TPU.

We first understand the impact of the systolic array size. \Fig{subfig:eval_bank} shows how the performance (FLOPS) and the array utilization changes when the array size increases from 32$\times$32 to 512$\times$512 when running the VGG model. 
As the array size increases, the performance increases while the utilization decreases. The utilization decreases by half when the array size increases from 128 to 256. This highlights the diminishing return of increasing the systolic array size, and corroborates the design decision of choosing a size of 128 for balancing peak FLOPS and utilization in TPUv2.

We also evaluate the choice of word size in the vector memory in TPU-v2. We use the OpenRAM SRAM compiler~\cite{guthaus2016openram} with the 45~nm (freepdk45) process to estimate the area overhead when changing the word size while fixing the SRAM capacity at 256~KB.

\Fig{fig:area} shows how the SRAM bandwidth idle ratio (in VGG16 inference) and the SRAM area change as the word size increases from 1 to 32. 

 The word size 8 achieves the area efficiency that is close to the minimum value, while the word size 1 leads to a $5 \times$ overhead.
As such, TPUv2 uses the word size 8 for the area efficiency.
However, with a word size of 8, the vector memory bandwidth utilization is below 50\%.
This insight explains why the TPUv3 chooses to add another systolic array to leverage this extra vector memory bandwidth~\cite{tpu_cacm}.

\begin{figure}[t]

    {
    \centering
      \includegraphics[trim=0 0 0 0, clip, width=0.8\linewidth]{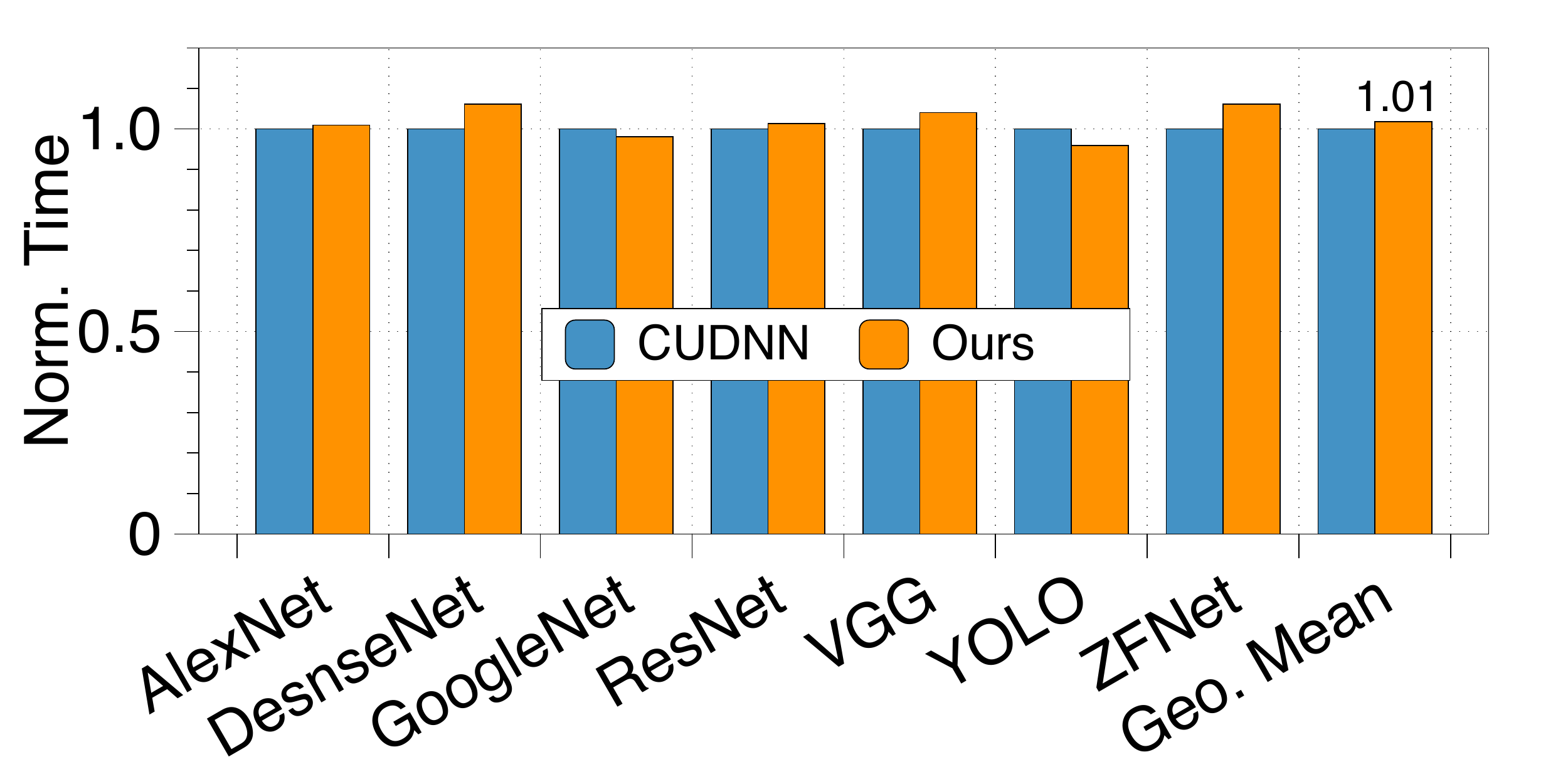}
      \label{subfig:bs8}
        \vspace{-0.2cm}

    }
    \vspace{-3pt}
    \caption{\small Normalized execution time of cuDNN and our implicit method on V100 GPU with tensor cores at batch size of 8.}
    \label{fig:GPU_time}
    \vspace{-0.4cm}
\end{figure}

\subsection{Evaluation on Tensor Cores}

\paragraph{End-to-end Model Results}
\Fig{fig:GPU_time} shows the execution time of our GPU implementation normalized to that of the baseline cuDNN implementation.
Our implementation is almost identical to the baseline, average 1\% slower when $N=8$ (\Fig{fig:GPU_time}).
We note that cuDNN uses low-level microarchitecture-specific optimizations~\cite{ma2020rammer} that are unavailable to us. The results indicate that our \im{} algorithm is competitive to Nvidia's proprietary implementation.

\begin{figure}[t]
    \vspace{-0.02cm}
	\centering
        \subfloat[Stride sensitivity.]
    {
      \includegraphics[trim=0 0 0 0, clip, width=0.49\linewidth]{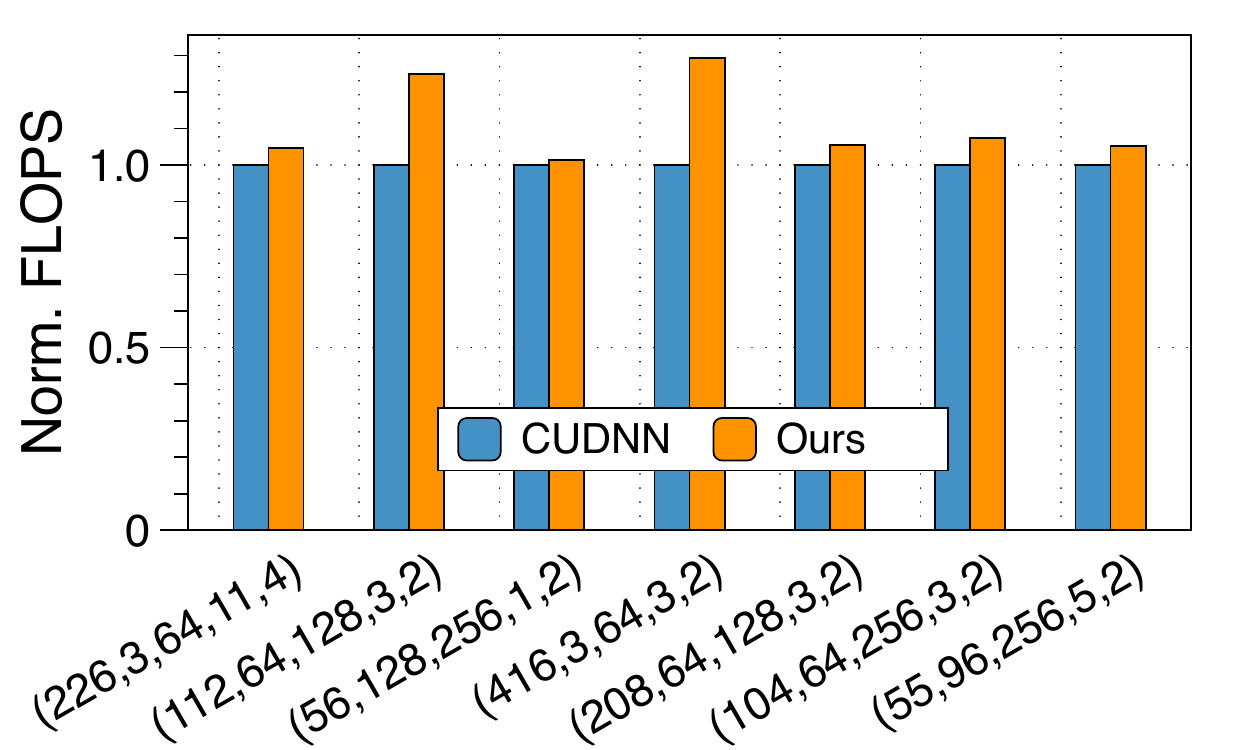}
      \label{subfig:gpu_model_stride}
  
    }%
    \subfloat[Inter-tile reuse impact.]
    {
        \includegraphics[width=0.49\linewidth]{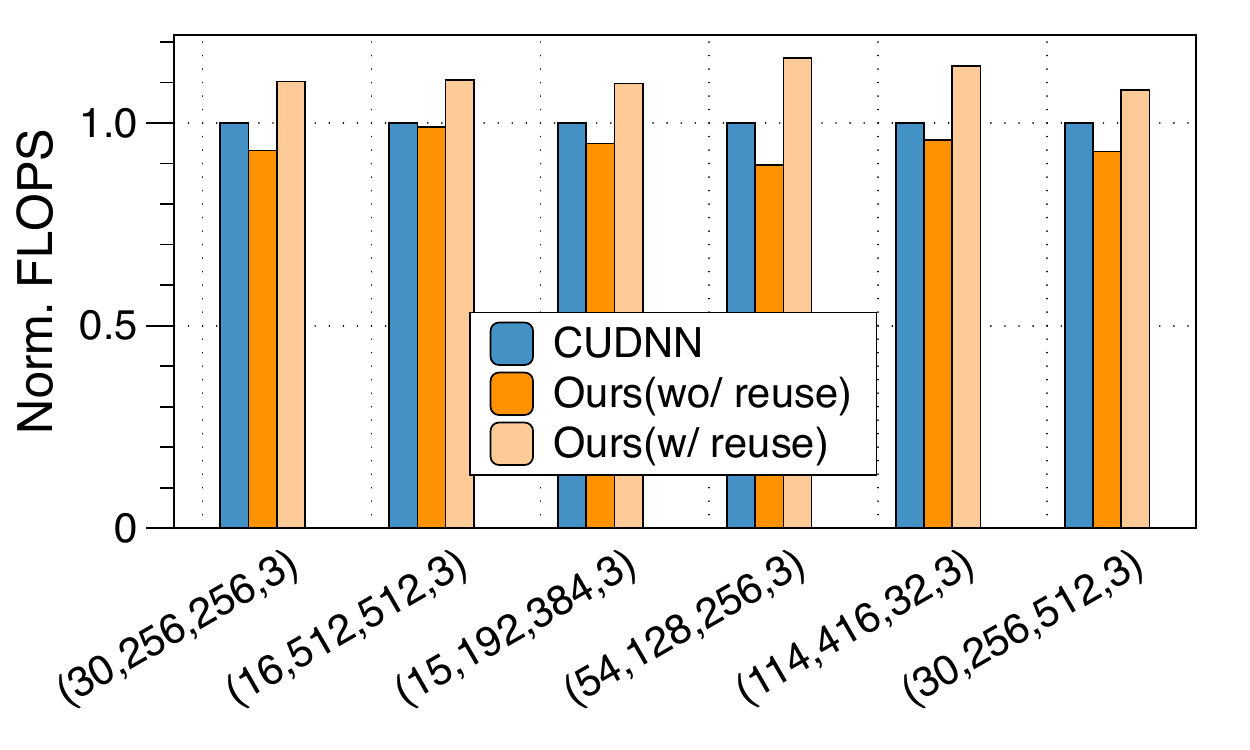}
        \label{subfig:shared_update}
    }
    \vspace{-3pt}
    \caption{\small Evaluation of our GPU optimizations. The digits of x-axis label indicate $W_I, C_I, C_O, W_F$, with an additional digit for stride in (a).}
    \label{fig:gpu_opt}
    \vspace{-0.4cm}
\end{figure}

\paragraph{Strided Convolution} The advantage of our method over the cuDNN is when the stride is greater than 1.
\Fig{subfig:gpu_model_stride} shows the performance (FLOPS) of our design normalized to that of cuDNN for \conv{} layers from the benchmarked models that have a stride value greater than one (indicated by the last digit of the $x$-axis label).
Our method is on average 20\%, up to 40\%, faster than cuDNN.

\paragraph{Inter-tile Reuse} \Fig{subfig:shared_update} evaluates the effectiveness of our inter-tile reuse optimization on GPUs for a set of layers in different models. These layers are used here because the overhead of global memory accesses in these layers is not completely overlapped by the computation in the pipeline, which makes it important to increase on-chip data reuse through inter-tile reuse.
Overall, our optimization leads to an average 16.7\% performance improvement.

\section{Related Work}
\label{sec:related}

\paragraph{im2col} Jia et al.~\cite{jia2014caffe} applies \im{} algorithm in a deep learning system to accelerate convolution operation in DNN inference and training. 
Some work~\cite{cho2017mec, 2017Low, anderson2017low} reduce memory overhead and improve the performance by modifying the lowered IFMap into a compact format, but still explicit GEMM.
Some studies~\cite{georganas2018anatomy, georganas2020harnessing} achieve efficient direct convolution through a special data layout to optimize computation.
Prior work Delta~\cite{lym2019delta} provides an analytical model for existing implicit channel-last \im{} on GPU, for which Duplo~\cite{kim2020duplo} provides a hardware-based acceleration solution.
In contrast, our work identifies the more general and flexible implicit channel-first \im{} that are likely used by TPUs.
We describe its implementation details on the systolic array architecture and software-level implementation on GPUs.

\paragraph{DNN Accelerators} Most of today's DNN accelerators, especially ones used in industry, target GEMM~\cite{tpu_paper, v100, medina2020habana, qin2020sigma, sma_dac20}, requiring some form of \im{}. Other accelerators target convolution~\cite{chen2014diannao, du2015shidiannao, zhang2015optimizing, Eyeriss} by designing a dedicated data flow for direct convolution. SCALE-Sim~\cite{samajdar2018scale} proposes a systolic array simulator accelerating GEMM and assumes an explicit \im{} execution method. 
Sparsity is an important property of DNN models, which prior works have exploited for acceleration~\cite{tw_sc20} and robustness enhancement~\cite{path_cvpr19,ptolemy_micro20}. 
As such, researchers have proposed various sparse accelerator designs, which are based on direct convolution ~\cite{cnvlutin, scnn} or assume the usage of explicit \im{}~\cite{sparten, zhu2019sparse, BitTactical} and the implicit channel-last \im{} algorithm~\cite{dualsparsetc}.

Our work identifies a generic \im{} algorithm that translates convolution to GEMM with practical zero-cost performance and memory overhead. We demonstrate that it can be applied to both TCs-like GEMM engines and systolic arrays. 
We believe that our work can encourage future study for designing sparse CNN accelerators based on the described channel-first implicit \im{} algorithm.

\section{Conclusion}
\label{sec:conclusion}

In this work, we propose an implicit \im{} algorithm called channel-first \im{} that is very likely used by TPUs.
It dynamically converts a convolution into a GEMM with zero performance and memory overhead. 
We describe its implementation details on the TPU architecture.
We also demonstrate its general applicability and performance advantages on other GEMM-engines such as the tensor cores on Nvidia's GPUs.
We hope this can encourage future work on accelerating CNNs on specialized GEMM engines. 

\vspace*{0.1cm}\noindent\textbf{Acknowledgements}\hspace*{0.2cm}
We thank the anonymous reviews for their thoughtful comments and suggestions.
Jingwen Leng is the corresponding author of this paper.
This work was supported by the National Key R\&D Program of China under Grant 2020YFB1711801, and National Natural Science Foundation of China (NSFC) grant 62072297.


{
\bibliographystyle{IEEEtran}
\bibliography{ref}
}

\end{document}